\documentclass[11pt,a4paper]{article}
\usepackage{amsmath, amsfonts,amssymb,mathtools,nicefrac,nccmath,cases}
\usepackage{microtype,booktabs,multirow}
\usepackage[usenames,dvipsnames,table]{xcolor}
\usepackage{colortbl}
\usepackage{tikz}
\usetikzlibrary{shapes,arrows}
\usetikzlibrary{calc,shapes.callouts,shapes.arrows,bending,intersections}
\usetikzlibrary{shapes.geometric}
\usetikzlibrary{arrows.meta,arrows}
\usetikzlibrary{decorations.pathmorphing}
\usetikzlibrary{decorations.markings}
\usetikzlibrary{shapes.multipart}
\usetikzlibrary{tikzmark}
\usepackage{arydshln}

\usepackage[nosort]{cite}
\usepackage{colortbl}

\usepackage{physics}

\usepackage{xcolor}
\usepackage{lipsum}

\usepackage[linktocpage=true]{hyperref}
\usepackage[noabbrev]{cleveref} 
\hypersetup{colorlinks=true,linkcolor=nicecolor,citecolor=nicecolor,urlcolor=nicecolor}
\definecolor{nicecolor}{rgb}{0.1, 0.3, 0.4}

\definecolor{blue}{rgb}{0.06, 0.3, 0.57}
\definecolor{Gray}{gray}{0.4}

\definecolor{nicecolor}{rgb}{0.1, 0.3, 0.4}
\definecolor{blue}{rgb}{0.06, 0.3, 0.57}
\definecolor{Gray}{gray}{0.4}
\colorlet{tableheadcolor}{gray!15} 
\colorlet{tablerowcolor}{gray!7} 

\def\hybrid{\topmargin -20pt    \oddsidemargin 0pt
	\headheight 0pt \headsep 0pt
	\textwidth 6.5in        
	\textheight 9in         
	\textwidth 6.25in       
	\textheight 9 in       
	\marginparwidth .875in
	\parskip 5pt plus 1pt 
	\jot = 1.5ex
}
\hybrid
\numberwithin{equation}{section}
\numberwithin{table}{section}
\usepackage{float}
\usepackage{tabularx}
\newcolumntype{D}{>{\centering\arraybackslash}X}
\newcolumntype{L}{>{$}l<{$}}
\newcolumntype{R}{>{$}r<{$}}
\newcolumntype{C}{>{$}c<{$}}
\usepackage{IEEEtrantools}
\usepackage{makecell}

\newcommand{\beq}{\begin{equation}}  \newcommand{\eeq}{\end{equation}}
\newcommand{\bal}{\begin{aligned}}   \newcommand{\eal}{\end{aligned}}
\newcommand{\bea}{\begin{eqnarray}}  \newcommand{\eea}{\end{eqnarray}}
\def\beqa{\begin{eqnarray}}
\def\eeqa{\end{eqnarray}}

\newcommand{\bmat}{\left(\begin{array}}
\newcommand{\emat}{\end{array}\right)}

\newcommand{\bbC}{\mathbb{C}}
\newcommand{\bbR}{\mathbb{R}}

\newcommand{\cO}{\mathcal{O}}

\newcommand{\cC}{\mathcal{C}}

\newcommand{\cN}{\mathcal{N}}
\newcommand{\cW}{\mathcal{W}}

\newcommand{\cR}{\mathcal{R}}

\newcommand{\cV}{\mathcal{V}}

\usepackage{cancel}

\newcommand{\be}{\begin{equation}}
\newcommand{\ee}{\end{equation}}

\newcommand{\im}{\mathbf{i}}
\newcommand{\slt}{\mathfrak{sl}(2)}
\newcommand{\SLt}{\mathrm{SL}(2)}
\newcommand{\rI}{\mathrm{I}}
\newcommand{\rII}{\mathrm{II}}

\newcommand{\rV}{\mathrm{V}}
\newcommand{\gR}{\mathfrak{g}_\bbR}

\newcommand{\bbZ}{\mathbb{Z}}
\newcommand{\subs}{\subset}
\newcommand{\spanR}[1]{\mathrm{span}_\bbR\{#1\}}

\newcommand{\conj}[1]{\overline{#1}}
\newcommand{\Hp}{\ensuremath{H_{\textrm{p}}}}
\newcommand{\Fpol}{\ensuremath{F_{\textrm{pol}}}}

\newcommand*\Bell{{\ensuremath{\boldsymbol\ell}}}

\newcommand{\hphi}{\ensuremath {\hat{\phi}}}

\newcommand{\svev}[1]{\ensuremath {\overline{s}^{#1}}}
\newcommand{\hN}{\ensuremath {N^{-}}}
\newcommand{\adY}[1]{\ensuremath {\mathrm{ad}_{N^{0}_{(#1)}}}}
\newcommand{\opY}[1]{\ensuremath {N^{0}_{(#1)}}}
\newcommand{\hf}{\ensuremath {\frac{1}{2}}}
\newcommand{\ua}{\ensuremath {\underline{\alpha}}}

\newcommand{\Ker}{\ensuremath {\mathrm{Ker\,}}}
\newcommand{\blank}{\ensuremath {{\,\cdot\,}}}
\newcommand{\ad}[1]{\ensuremath {\operatorname{ad}_{#1}}}
\newcommand{\fg}{\ensuremath {\mathfrak{g}}}
\newcommand{\hg}{\ensuremath {\hat{g}}}

\definecolor{Gray}{gray}{0.95}

\begin{document}

\baselineskip=14pt
\parskip 5pt plus 1pt

\vspace*{-1.5cm}
\begin{flushright}    
  {\small 
  
  }
\end{flushright}

\vspace{2cm}
\begin{center}        

  {\huge Universal Axion Backreaction\\[.3cm]
  in Flux Compactifications}
  
\end{center}

\vspace{0.5cm}
\begin{center}        
{\large  Thomas W.~Grimm\footnote{t.w.grimm@uu.nl} and Chongchuo Li\footnote{c.li@uu.nl}}
\end{center}

\vspace{0.15cm}
\begin{center}        
\emph{ Institute for Theoretical Physics \\
Utrecht University, Princetonplein 5, 3584 CE Utrecht, The Netherlands}
 
\end{center}

\vspace{2cm}


\begin{abstract}
\noindent
We study the backreaction effect of a large axion field excursion on the saxion partner residing in the same 
$\cN=1$ multiplet. Such configurations are relevant in attempts to realize axion monodromy inflation in string compactifications. 
We work in the complex structure moduli sector of Calabi-Yau fourfold compactifications of F-theory with four-form fluxes, which 
covers many of the known Type II orientifold flux compactifications. Noting that 
axions can only arise near the boundary of the moduli space, the powerful results of asymptotic Hodge theory provide 
an ideal set of tools to draw general conclusions without the need to focus on specific geometric examples. 
We find that the boundary structure engraves a remarkable pattern in all possible scalar potentials generated by background fluxes. 
By studying the Newton polygons of the extremization conditions of all allowed scalar potentials and realizing the backreaction effects as Puiseux expansions, we find that this pattern forces a universal backreaction behavior of the large axion field on its 
saxion partner.
\end{abstract}

\thispagestyle{empty}
\clearpage

\setcounter{page}{1}


\newpage

\begingroup
  \flushbottom
  \tableofcontents
  \newpage
\endgroup


\section{Introduction and discussion}
\label{sec:intro}

Axion monodromy inflation \cite{Silverstein:2008sg,McAllister:2008hb,Kaloper:2008fb,Marchesano:2014mla,McAllister:2014mpa} is an intriguing suggestion to realize large-field inflation in string theory. In such models, an axion is initially placed at a transplanckian distance away from its true vacuum and then rolls down towards the vacuum to drive inflation. 
The naive axion periodicity is extended in axion monodromy models
by first unfolding the field range of the axion to the real line, i.e.~by extending its field range to the universal covering space 
of the periodic axion space. An axion scalar potential is then introduced to break the approximate continuous shift symmetry of the axion. 
Such a potential is required to fulfil a modified realization of a discrete axionic shift-symmetry in the sense that it is invariant under a combined transformation of the axion and some of the parameters in the potential, e.g.~flux numbers. The combined transformation is called a monodromy
transformation. In potentials realizing such monodromy symmetries one can then hope to implement inflation over multiple periods of the axion in a controlled fashion. 

Early studies \cite{Silverstein:2008sg,McAllister:2008hb} of axion monodromy inflation focus on its realization in non-supersymmetric Type II string compactifications with various NS-branes or D-branes. To establish such compactifications delicate control issues have to be addressed
to ensure the validity of the reduction. To have from the outset more 
control over the stability of the setting, so-called F-term axion monodromy models were proposed in  \cite{Marchesano:2014mla,Blumenhagen:2014gta,Hebecker:2014eua,Ibanez:2014kia,McAllister:2014mpa,Bielleman:2016olv}. These 
models consider axion monodromy inflation in the context of supersymmetric flux compactifications of string theory and will serve as the main motivation for this work. In these settings the axion scalar potential is induced by turning on various background fluxes of higher form fields and the monodromy is realized by simultaneously shifting the axion and some flux numbers. An apparent merit of the F-term axion monodromy inflation is that 
various moduli of the compactified string theory share the flux-induced F-term scalar potential with the axion inflaton candidates, so that the study of the moduli stabilization and axion inflation are naturally linked. 

The connection between the axion and moduli scalar potential forces one to consider the backreaction of the axion inflaton candidate on the vacuum expectation value of the moduli fixed by the F-term scalar potential. Since large field inflation requires a transplanckian displacement of the axion, 
it is alarming when such a large-field backreacts on the vacuum expectation value of geometric moduli of the compactification, since this 
requires to consider the dynamics of the combined axion-saxion system. This issue is pointed out in \cite{Baume:2016psm} and further analysed in \cite{Valenzuela:2016yny, Blumenhagen:2017cxt}, by investigating examples in various classes of string compactifications. Their findings confirm the worry, that in these examples, a large axion excursion does indeed backreact on the geometric moduli in a linear fashion, so that the moduli is pushed towards the boundary of the moduli space at infinite distance. Invoking the  Distance Conjecture \cite{Ooguri:2006in, Klaewer:2016kiy} the axion monodromy inflation program then faces more challenges as studied in the examples of \cite{Baume:2016psm,Valenzuela:2016yny,Blumenhagen:2017cxt}. More precisely, the Distance Conjecture states that in an effective field theory consistent with quantum gravity, a transplanckian displacement in field space is accompanied by additional light degrees of freedom, signalling a breakdown of the effective field theory. The linear backreaction of the axion onto the saxion can thus be responsible for a breakdown of 
the effective theory. It remains an open debate \cite{Blumenhagen:2014nba,Hebecker:2014kva,Baume:2016psm,Valenzuela:2016yny,Bielleman:2016olv,Blumenhagen:2017cxt,Landete:2017amp,Kim:2018vgz,Buratti:2018xjt} whether or not this backreaction can be avoided or sufficiently delayed to realize F-term axion monodromy models, see also the recent \cite{Calderon-Infante:2020dhm} where the authors argue that the linear backreaction is the maximum non-geodesity that would be consistent with the Distance Conjecture from a bottom-up perspective. The arising problems are also related to the difficulty of creating mass hierarchies in flux compactifications, as studied in \cite{Blumenhagen:2014nba,Hebecker:2014kva}.

In order to study the axion backreaction problem, it is desirable to identify a general and controlled setting with a variety of axion-like fields. The complex structure moduli space of Calabi-Yau manifolds provides such a general arena. Firstly, it has long been known that the complex structure moduli descent to complex scalars in the effective theory. Secondly, it became clear in the advent of flux compactifications that these fields can 
obtain a scalar potential when allowing for non-trivial background fluxes \cite{Grana:2005jc,Douglas:2006es}. 
Within such flux compactifications, it was found in \cite{Garcia-Etxebarria:2014wla,Grimm:2019ixq} that fields with approximate continuous shift symmetry, i.e.~axion fields, can only arise near certain boundary components of the complex structure moduli space. Near such a boundary component, the monodromy as one loops around the boundary descents to an approximate continuous shift-symmetry of the real part of the considered complex structure modulus, which is broken into a genuine discrete axionic shift-symmetry by non-perturbative effects. Moreover, by turning on background fluxes, one gets a scalar potential for the complex structure moduli which are decomposed into the axionic-like fields and the moduli controlling the geometry, and it is the very same monodromy transformation generating the shift-symmetry that is used as the monodromy transformation in the axion inflation scenario. 

A concrete setting that allows us to examine complex structure axions and their flux-induced scalar potentials 
arises from F-theory compactifications 
on Calabi-Yau fourfolds. Such compactifications lead to an $\cN=1$ supersymmetric effective action \cite{Denef:2008wq,Grimm:2010ks} with a classical scalar potential induced by background four-form flux $G_4$. It is also well-known that a part of $G_4$ induces a non-trivial F-term potential \cite{Gukov:1999ya} and it will be this part which will be relevant in the present work. 
The merit of studying complex structure axions in F-theory compactifications is at least two-fold. Firstly, the complex structure axions include the 
R-R zero-form axions of Type IIB string theory and also are dual to NS-NS two-form axions of Type IIA via duality. Moreover, one consistently 
incorporates certain axions that are associated to D-branes. Secondly, the scalar potential induced by the four-form flux in F-theory is particularly amenable to be studied by asymptotic Hodge theory \cite{MR0382272, MR840721} near the boundary of the complex structure moduli space
as demonstrated in \cite{Grimm:2019ixq,Grimm:2020cda}. These techniques have developed into a powerful tool to address various 
swampland conjectures in an example independent way \cite{Grimm:2018ohb,Grimm:2018cpv,Corvilain:2018lgw,Grimm:2019bey,Grimm:2019ixq,Gendler:2020dfp,Lanza:2020qmt,Grimm:2020cda,Bastian:2020egp}. In particular, it has been observed in \cite{Grimm:2019ixq} that one can systematically analyze the 
axion backreaction using asymptotic Hodge theory.
 
 In this work 
 we generalize and complete the analysis of \cite{Grimm:2019ixq} and investigate, in full generality, the backreaction within 
 a single axion-saxion pair arising near any boundary 
 of the complex structure moduli space. Remarkably, we are now able to combine the insights from 
asymptotic Hodge theory and Newton polygons associated with Puiseux expansions to establish that there is universally a 
backreaction of the axion on its saxion partner which grows when considering large field values of the axion. 
To achieve this goal we first note that the asymptotic Hodge theory provides near every boundary in moduli space an approximation to the all 
relevant functions in the effective theory via the so-called nilpotent orbit \cite{MR0382272}. The nilpotent orbit thus gives a 
near-boundary expansion that, by using the results of \cite{MR0382272,MR840721}, can be encoded by a set of boundary data.\footnote{It was recently suggested in \cite{Grimm:2020cda} that this is reminiscent of a holographic perspective, with an 
actual underlying bulk and boundary theory.}   A crucial part of this boundary data is a set of commuting $\slt$-algebras which decompose the middle cohomology of Calabi-Yau manifold into different $\slt$-representations. Also splitting a general four-form flux into such  
representations gives us precise control about the limiting behavior of Hodge norm and allows us to make 
the axion and saxion dependence in the scalar potential explicit. In fact, we find that the underlying $\slt$-structure ensures that the possible asymptotic scalar potentials generated by four-form fluxes form a rather constrained set. Systematically going through all allowed cases we can then 
study the axion backreaction generally. We do this by explicitly deriving the saxion vacuum expectation 
value determined by the extremization equation of the scalar potential. We evaluate this saxion vacuum value 
depending on the axion as a parameter and study the behavior of the solution in the limit when this parameter becomes large. The solution to this problem is given by Puiseux expansions, which present the solution as a fractional power series whose leading power is determined by a handy graphical tool called the Newton polygon. With the scalar potentials derived using asymptotic Hodge theory the Newton polygon provides visual guidance to the leading backreaction behavior of the saxion vacuum expectation value. This lets us uncover a universal backreaction behavior generalizing the result of \cite{Baume:2016psm,Valenzuela:2016yny,Blumenhagen:2017cxt,Grimm:2019ixq} that is present around any singularity in the Calabi-Yau fourfold complex structure moduli space.

Now we would like to state our results more concretely. In this work we focus on the backreaction of one complex structure axion field $\phi$ on the saxion modulus $s$ in the same $\cN=1$ multiplet. We find that, when a vacuum exists, a large displacement of the axion $\phi$ always backreacts on the saxion vacuum expectation value $\svev{}$ in the following fashion
\begin{equation} \label{scaling_intro}
  \svev{}(\phi) = c \phi^\gamma + \cO\bigg(\frac{1}{\phi}\bigg)\, ,
\end{equation}
where the prefactor $c > 0$ is a positive number and the exponent $0 < \gamma \le 2$ is a rational number. There could be several different sets of $c$ and $\gamma$ corresponding to one flux configuration. We find that in almost every case, one has a solution with $\gamma = 1$, which agrees with the linear backreaction behavior found in \cite{Baume:2016psm,Valenzuela:2016yny,Blumenhagen:2017cxt,Grimm:2019ixq}. The cases with $\gamma > 1$ are rather restricted to the extent that only $\gamma = 2$ is allowed and they also need to satisfy a technical condition which is discussed at the end of section~\ref{saxion_vacuum}. We find only two flux configurations that generate $\gamma = 2$ backreaction and they have no $\gamma = 1$ branch. In contrast, we do not find additional restrictions on cases with 
$0 < \gamma < 1$ besides that the exponent $\gamma$ should be rational. However, we note that for these cases there is sometimes a co-existing $\gamma = 1$ branch. Regarding the prefactor $c$ in the backreaction, the possibility that $c$ depends on flux numbers cannot be ruled out. Physically this implies that one cannot completely exclude a delayed backreaction. In section~\ref{saxion_vacuum} we give a simple condition on the flux such that the delay cannot occur. 

In view of our findings there are several interesting further studies that can be undertaken. Recalling the relation with the Distance Conjecture, it would be interesting to study the $\gamma \ne 1$ cases and the condition on the delay of a backreaction in the future. One first extension is to take into account subleading corrections following the strategy in \cite{Grimm:2020cda} and check whether these extend the class of scalar potentials can arise. Furthermore, in this work we have only exploited the conditions for $ \svev{}$ to be 
at an extremum of the potential and not incorporated the additional constraints that actually is at a minimum. Hence, the relation \eqref{scaling_intro} gives the necessary behavior also at minima, but it could well be that some of the cases with $\gamma \neq 1$ are not arising in an actual minimum. 
We would also like to point out that while for finding extrema it is sufficient to focus on one axion-saxion pair, the analysis of the conditions for having 
an actual minimum requires a more complete treatment of all involved moduli. We have already formulate the setting and the analysis in a general multi-variable language and we believe that it is desirable, while technically more difficult, to generalize the study of backreaction to a multi-variable setting where the vacuum expectation values of all saxions are considered.

This paper is structured as follows. In section \ref{sec:fOnCY4WithG4Flux}, we introduce the physical setting of Calabi-Yau fourfold compactifications of F-theory, review variations of Hodge structures, and rewrite the F-theory scalar potential using Hodge theory. In section \ref{sec:complexStructureAxionsAndAsympHodgeTheory}, we discuss the reason why only near the boundary of the complex structure moduli space axions can emerge, review the relevant part of asymptotic Hodge theory, and provide the asymptotic form of the F-theory scalar potential near the boundary. In section \ref{sec:axionDependentPotential}, after briefly setting up the notation for $\slt$-representations and reviewing the notion of Puiseux series and its associated Newton polygon, we expand the asymptotic scalar potential and solve the backreacted vacuum expectation value of the geometric moduli. There we will not only encounter the general backreaction behavior, but also collect apparent counterexamples. Those counterexamples are then ruled out at the end of that section. Finally, in appendix \ref{app:commutingsl2} we provide some technical details of asymptotic Hodge theory used in the main text.

\section{F-theory on Calabi-Yau fourfolds with $G_4$-flux} \label{sec:fOnCY4WithG4Flux}

In this section we introduce in more detail the context in which we study axion backreaction and moduli stabilization. More precisely, we will 
introduce part of the  low-energy supergravity theory arising when considering F-theory compactified on a family of Calabi-Yau fourfolds $Y$ carrying $G_4$-flux in section \ref{scal1}. We will then express the induced flux scalar potential depending in terms 
of the Hodge decomposition in section \ref{Hodge-filtr} and comment on the use of the Hodge filtration.  

\subsection{Scalar potential and its complex structure dependence} \label{scal1}

To begin with, we first compactify M-theory with $G_4$-flux on Calabi-Yau fourfolds $Y$ to obtain a three-dimensional $\cN = 2$ effective supergravity theory with a
scalar potential for the complex structure moduli and K\"ahler structure moduli induced by the flux \cite{Haack:2001jz}. To connect this setting to an F-theory compactification we assume that $Y$ admits a two-torus fibration. The three-dimensional action is then lifted to a four-dimensional $\cN = 1$ supergravity theory by shrinking the volume of the two-torus fiber of $Y$. This procedure defines the reduction of F-theory on the Calabi-Yau fourfold $Y$ to obtain a four-dimensional effective theory \cite{Denef:2008wq,Grimm:2010ks}. Crucial for our considerations is the fact that 
the complex structure moduli of $Y$ reside in chiral multiplets both in the three-dimensional effective theory obtained from M-theory and the four-dimensional effective theory derived by the lift to F-theory. It was shown in \cite{Garcia-Etxebarria:2014wla,Grimm:2018ohb,Grimm:2019ixq} 
that within the complex structure moduli 
space, fields with approximate shift symmetry, i.e.~axion fields, can only arise near the boundaries of moduli space. These boundaries 
will have to satisfy certain conditions, which we will recall below. It is with these axions and their partner saxions that will be the focus of this work. 

Let us now discuss the scalar potential induced by a background flux
 $G_4 \in H^4(Y, \bbZ/2)$. It is well-known \cite{Haack:2001jz} that the three-dimensional scalar potential arising in the M-theory 
 reduction takes the form
\begin{equation} \label{V-M}
  V= \frac{1}{\cV_4^3}\left(\int_Y G_4 \wedge \ast G_4 - \int_Y G_4 \wedge G_4\right)\, ,
\end{equation}
where $\ast$ is the Hodge-star operator of $Y$ and $\cV_4$ is the volume of the Calabi-Yau fourfold $Y$. Note that the scalar potential depends on 
complex structure and the K\"ahler structure deformations through the Hodge-star operator in the first integral. 
Furthermore, in $V$ there is an additional K\"ahler structure moduli dependence through the overall volume factor. Without further inclusion of localized sources such as M2-branes filling the three-dimensional spacetime, the $G_4$-flux needs to satisfy the following tadpole cancellation condition \cite{Sethi:1996es}
\begin{equation}
  \frac{1}{2} \int_Y G_4 \wedge G_4 = \frac{\chi(Y)}{24}\, .
\end{equation}

The scalar potential \eqref{V-M} can be cast into a form compatible with $\cN=2$ supersymmetry in three dimensions. Instead 
of reviewing the whole construction of the characteristic $\cN=2$ data, we will henceforth focus only 
on the complex structure moduli dependence of the Hodge star in \eqref{V-M}. This amounts to requiring that our 
$G_4$-flux lives in the \emph{primitive} middle cohomology $\Hp^4(Y, \bbZ)$ \cite{Haack:2001jz}, which is equivalent to stating  
that 
if $J \in H^{1, 1}(Y)$ is the K\"ahler class of the Calabi-Yau fourfold $Y$ the fluxes under consideration satisfy 
the condition $J \wedge G_4 = 0$. Inserting this condition into \eqref{V-M}, the resulting scalar 
potential can be shown to arise from a K\"ahler potential $K$ and a superpotential $W$ \cite{Gukov:1999ya,Haack:2001jz} given by 
\beq \label{KW-cs}
   K = - \log \int_Y \Omega \wedge \bar \Omega\, , \qquad     W = \int_Y \Omega \wedge G_4\, , 
\eeq
where $\Omega$ is the, up to rescaling, unique $(4,0)$-form on $Y$ and hence represents $H^{4,0}(Y, \bbC)$. 

Let us note that we make a further simplification that will keep our discussion accessible. In order to not 
deal with integer cohomology and the many associated subtleties, we will not be careful about the 
quantization of the $G_4$ flux \cite{Witten:1996md} and ofter write $G_4 \in \Hp^4(Y, \bbR)$. While 
imposing the quantization conditions on $G_4$ is important in deriving numerical values and establishing finiteness results \cite{Grimm:2020cda,GrimmSchnellinprogress}
it will not be altering the conclusions about backreaction that we obtain in this work. 

\subsection{Scalar potential and the Hodge filtration} \label{Hodge-filtr}

In order to identify the conditions for axions to arise in the complex structure moduli space and to study their backreaction effects, 
we need to put the above expression for the scalar potential into a Hodge-theoretic context. Let us 
review some of the relevant definitions in order to establish notations. For more complete information, we refer the math article \cite{MR1042802}
and the recent physics application \cite{Grimm:2018cpv,Grimm:2020cda}.

To begin with, we note that the primitive middle cohomology of a smooth Calabi-Yau fourfold $Y$ carries a pure Hodge structure of weight four, which is described by the Hodge decomposition
\begin{equation} \label{primHodge-dec}
  \Hp^4(Y, \bbC) = H^{4, 0} \oplus H^{3, 1} \oplus H^{2, 2} \oplus H^{1, 3} \oplus H^{0, 4}\, ,
\end{equation}
satisfying $H^{p, q} = \conj{H^{q, p}}$. Note that the $H^{2, 2}$ here denotes \emph{primitive} $(2, 2)$-forms, i.e.~it is a shorthand for $H^{2, 2} \cap \Hp^4(Y, \bbC)$, and all subspaces $H^{p, q}$ different from $H^{2, 2}$ are automatically primitive in the Calabi-Yau fourfold setting.
In practice it is more useful to use an equivalent description in terms of the \emph{Hodge filtration}
\begin{equation}
  0 \subs F^4 \subs F^3 \subs F^2 \subs F^1 \subs F^0 = \Hp^4(Y, \bbC)\, ,
\end{equation}
satisfying $F^p \oplus \conj{F^{4 - p + 1}} \cong \Hp^4(Y, \bbC)$. One can recover the Hodge decomposition by setting $H^{p, q} = F^p \cap \conj{F^q}$. On the other hand, given a Hodge decomposition, the corresponding Hodge filtration is given by 
\beq \label{Fp-intro}
    F^p = \bigoplus_{r \ge p} H^{4 - r, r}\, .
\eeq
The Hodge structure will change as one deforms the complex structure of the Calabi-Yau, and the merit of using Hodge filtration is that the filtration will change \emph{holomorphically} with respect to the complex structure moduli $t^I$.  The variation of Hodge structure is captured by the period map $F(t)$, which records the whole Hodge filtration $F^p$, $p=0,...,4$, corresponding to the complex structure moduli $t^I$.

Moreover, the Hodge structure is polarized by the symmetric intersection form on the Calabi-Yau fourfold
\begin{equation}
  \langle\alpha, \beta\rangle = \int_Y \alpha \wedge \beta\, ,
\end{equation}
for $\alpha, \beta \in \Hp^4(Y, \bbC)$. There is also a \emph{Weil operator} $C_F$, depending on the Hodge filtration $F$, defined by
\begin{equation} \label{eqn:defnWeilOperator}
  C_F(\alpha) = \im^{p - q} \alpha\, , \quad \textrm{for } \alpha \in H^{p, q}\, .
\end{equation}
It is a standard result that on the primitive middle cohomology of a Calabi-Yau fourfold, the Weil operator coincides with the Hodge star operator. With the help of the polarization form and the Weil operator, we can define the \emph{Hodge inner product} and its associated \emph{Hodge norm}
\begin{align}
  \braket{\alpha}{\beta}_F & = \langle C_F \alpha, \conj{\beta} \rangle\, ,\label{eqn:defnHodgeInnerProduct}\\
  \Vert \alpha \Vert^2_F   & = \braket{\alpha}{\alpha}_{F}\, ,\label{eqn:defnHodgeNorm}
\end{align}
which depends on the Hodge filtration $F$.

For later reference, we also need to introduce the symmetry group of the variation of Hodge structures: It is the group $G$ of linear automorphisms of $\Hp^4(Y, \bbC)$ that preserves the dimension of Hodge filtration and the polarization pairing $\langle \blank, \blank \rangle$. There is also a real counterpart $G_\bbR$ of this symmetry group that consists of the automorphisms of $\Hp^4(Y, \bbR)$. More concretely, in the case of Calabi-Yau fourfolds, we have
\begin{equation} \label{groups}
  G = \mathrm{SO}(2 + h^{2, 2}_{\rm p} + 2h^{1, 3}, \bbC)\, , \quad \text{ and } \quad G_\bbR = \mathrm{SO}(2 + h^{2,2}_{\rm p} , 2h^{1, 3})\, ,
\end{equation}
where $h^{2, 2}_{\rm p} = h^{2, 2} - h^{1, 1}$ is the dimension of the space of primitive $(2,2)$-forms. This number follows from the Lefschetz decomposition. Namely, one has
\begin{equation}
  H^{2, 2} = \Hp^{2, 2} \oplus J \Hp^{1, 1} \oplus J^2 \Hp^{0, 0}\, ,\quad H^{1, 1} = \Hp^{1, 1} \oplus J \Hp^{0, 0}\, ,\quad \Hp^{0, 0} = H^{0, 0}\, ,
\end{equation}
where $J H^{p, p} = \{J \wedge \alpha | \alpha \in H^{p, p}\} \subs H^{p + 1, p + 1}$ contains the cup product between the K\"ahler class and $(p, p)$-classes. Since we are working with Calabi-Yau spaces, $h^{0, 0} = 1$. From the last two equalities, one finds $h^{1, 1}_{\textrm{p}} = h^{1, 1} - 1$. And then $h^{2, 2}_{\textrm{p}} = h^{2, 2} - h^{1, 1}$ follows.

We can now re-express the $F$-term scalar potential in the Hodge theoretical language: Note that our $G_4$ is real, so the scalar potential induced by $G_4$ can now be written as
\begin{equation} \label{scalar-potential}
  V(t) = \frac{1}{\cV_4^3}(\Vert G_4 \Vert^2_{F(t)} - \langle G_4, G_4 \rangle)\, .
\end{equation}
The first term contains the Hodge norm of $G_4$ evaluated at the Hodge structure $F(t)$. This is the term that we will focus on in the study of the backreaction of axions on the vacuum expectation values of stabilized saxions.

\section{Complex structure axions and asymptotic Hodge theory} \label{sec:complexStructureAxionsAndAsympHodgeTheory}

In this section we explain that in order to identify axionic directions in complex structure moduli space we
have to be at its boundaries, i.e.~approach a limit in which the associated Calabi-Yau fourfold degenerates.
Furthermore, we will see that the considered boundary has to satisfy a set of conditions \cite{Garcia-Etxebarria:2014wla,Grimm:2018ohb,Grimm:2019ixq}. 
To formulate these conditions we have to introduce some additional facts about the moduli space and, in 
particular, the behavior of the Hodge decomposition \eqref{primHodge-dec} near its boundaries. 
This forces us to briefly review parts of asymptotic Hodge theory that is relevant to our study 
of the axion backreaction problem. 
In section \ref{sec:whyaxions} we will briefly recall why axions arise near the boundaries of the moduli 
space and introduce the so-called nilpotent orbit that describes the asymptotic form of the Hodge decomposition. 
In section \ref{sec:commSl2} we then explain how one can associate to each boundary sets of commuting $\slt$-triples and a well-defined boundary 
Hodge structure. The asymptotic form of the scalar potential is then determined in section \ref{sec:normalFormPeriodMapping}. We first 
introduce the normal form of the period mapping $F(t)$ near any boundary and use it to find the asymptotic expression of the Hodge norm. That asymptotic expression will be used in our study of the axion backreaction problem in the next section. The aim of this section is to briefly introduce the relevant results without going into mathematical details. We will supply further details in appendix \ref{app:commutingsl2}. The precise mathematical statements summarized in this section are contained in the review \cite{MR1042802} and the original papers \cite{MR0382272}, \cite{MR840721}. For a more physical formulation that emphasizes similar parts of asymptotic Hodge theory, see \cite{Grimm:2018cpv,Grimm:2020cda}.

\subsection{Axions at the boundary of moduli space} \label{sec:whyaxions}

In this section we identify the regions in the complex structure moduli space in which one can 
find fields that admit approximate continuous 
shift symmetries and hence can be interpreted as axions. In order to do that we 
first recall that the discrete symmetries of the moduli space are encoded by the so-called monodromy 
group. These monodromy symmetries 
can then lead to an approximate continuous shift symmetry near the boundaries of moduli space. 

In the following we will introduce stepwise a local description of the region near the boundaries in 
moduli space along which the associated Calabi-Yau manifold $Y$ degenerates. We 
begin by fixing local coordinates $z^I$, where $I = 1, \ldots, h^{3, 1}$, 
for a patch around a boundary locus of codimension $n$ in the complex structure moduli space. 
We choose these coordinates such that the boundary is located at $z^i = 0$ for all $i = 1, \ldots, n$. 
It will also often be useful to implement the coordinate transformation \footnote{The coordinates $t^i$ are actually 
local coordinates on the universal cover of the near boundary patch in the moduli space.}
\begin{equation} \label{eqn:defnOfti}
	t^i = \frac{1}{2\pi\im}\log z^i\, .
\end{equation}
The new set of coordinates $t^i$ take value in the upper half plane, and the singularity is now located at $t^i \to \im\infty$, for $i = 1, \ldots, n$. The coordinates $t^i$ can be further decomposed into real and imaginary parts
\begin{equation}
  t^i = \phi^i + \im s^i\, .
\end{equation}
The  real parts $\phi^i$ are the candidate axions if one is close to the boundary $s^i = \infty$ as we will 
see below. In fact, the $\phi^i$ can enjoy an approximate shift-symmetry if the monodromy 
transformation associated to the boundary satisfies certain conditions. In cases in which 
the $\phi^i$ are identified as axions the imaginary parts $s^i$ is often referred to as saxion. We will sometimes 
use this terminology more loosely, by referring to the coordinates $s^i$, $\phi^i$ as saxion and axion without always 
stressing the extra condition on the associated monodromy transformation.  
Note that from \eqref{eqn:defnOfti} we see that $t^i$ takes value in the upper half plane, so we have a 
basic constraint $s^i > 0$. The region close to the boundary is characterized by $s^i \gg 1$. 

The next data one needs to record is the monodromy operators that arise when encircling the 
boundary locus $z^i=0$, $i=1,...,n$. There are $n$ monodromy operators and they are defined as the monodromy of the period map $F(t)$, introduced after \eqref{Fp-intro}, when one loops around the singular locus: When $t^i \to t^i + 1$ (equivalently $z^i \to e^{2\pi\im}z^i$), the period map changes $F(t^i + 1) = T_i F(t^i)$. In our geometric setting, where the variation of Hodge structure is induced from the deformation of Calabi-Yau complex structures and that there is an integral basis of the primitive middle cohomology $\Hp^4(Y, \bbC)$, one can choose the coordinates in the complex structure moduli space such that the monodromy matrices take the form
\begin{equation} \label{eqn:defnLogMonodromy}
  T_i = e^{N_i}\, ,
\end{equation}
where $N_i$ are nilpotent matrices. Note that the operators $T_i$ are elements of the real symmetry group $G_\bbR$ given in \eqref{groups}, 
while the $N_i$ are elements of the associated real algebra $\fg_\bbR$.

The monodromy matrices $T_i$, or rather the associated $N_i$, are essential in evaluating the Hodge decomposition \eqref{primHodge-dec} near the 
boundary. In general, the Hodge structure will degenerate exactly on the boundary $s^i = \infty$ and has to be replaced by a more sophisticated 
structure, a so-called `limiting mixed Hodge structure', which we will describe in section \ref{sec:commSl2}.
The starting point for construction of this structure is Schmid's \emph{nilpotent orbit theorem}\cite{MR0382272,MR840721}. 
It states that around the boundary locus $s^i = \infty$, i.e.~when $s^i \gg 1$, the period map $F(t)$ 
is well approximated by the \emph{nilpotent orbit} of the following form
\begin{equation} \label{eqn:nilpotentOrbit}
  \Fpol(t) = e^{t^i N_i} F_0\, ,
\end{equation}
where we sum in the exponential over $i=1,\ldots, n$.
In fact, the nilpotent orbit can be viewed as the essential part of the period map that arises by dropping certain exponential corrections 
$O(e^{2\pi i t^j})$,
while still keeping a well-defined Hodge decomposition. Note 
that this implies that in many limits non-perturbative corrections are still recorded in $\Fpol(t) $ as discussed in more detail in \cite{Grimm:2020cda, Bastian:2021eom}. Note that \eqref{eqn:nilpotentOrbit} immediately implies 
that in order that $\Re t^i = \phi^i$ is an axion with an approximate shift symmetry unbroken by 
$O(e^{2\pi i t^j})$-corrections we 
have to consider a boundary with $N_i$ non-vanishing.

While we will bypass using the K\"ahler potential and superpotential \eqref{KW-cs}, let us remark that we have given these quantities in terms of the holomorphic $(4,0)$-form $\Omega$. In order to get the naive nilpotent orbit approximation $K_{\rm nil}$
for $K$ we can now apply the fact that also $F^4 = H^{4,0}$ admits 
a representation  \eqref{eqn:nilpotentOrbit}. This implies that $\Omega$, after possibly fixing an overall rescaling, 
can be expressed as 
\beq \label{Pinil}
     \Omega (z)   =  \underbrace{e^{t^i N_i} a_0 }_{ \Omega_{\rm nil} ( t)} + \cO(e^{2\pi i t^j})\, .
\eeq
Here 
$a_0$ can still be a holomorphic function in the coordinates that are not sent to a limit. 
 Inserting this expression we can now write 
\beq \label{Knil}
  K_{\rm nil} =   - \log \langle \Omega_{\rm nil} , \bar{\Omega}_{\rm nil} \rangle = - \log \langle e^{2i s^j N_j}a_0  ,\bar{a}_0  \rangle \, ,
\eeq
which is thus a logarithm of a polynomial in the $s^i$ with a finite number of terms.
This implies that $K_{\rm nil}$ is independent of the axions $\phi^i$, while 
$K_{\rm nil}$ still depends on a considered variable $s^i$ if $N_i a_0 \neq 0$. 
We therefore conclude that a sufficient condition that $\phi^i$ is an axion with an 
approximate continuous shift symmetry $\phi^i \rightarrow \phi^i + c^i$ is that $N^i a_0 \neq 0$. 
This latter condition is a necessary condition for the limit 
to be at infinite distance in the metric derived from $K$ \cite{MR1432818}.
The appearance of the continuous shift symmetries at infinite distance singularities
was discussed in \cite{Grimm:2018ohb} in the context of the  Distance Conjecture. In the 
following, we will not restrict our attention to cases where $N_i a_0 \neq 0$, since 
this restriction is not necessary for our arguments to go through. While we have 
not checked that the K\"ahler metric indeed depends on the $\phi^i$ only through exponential 
corrections, we will see that in either case this does not alter our analysis. In the following we
will refer to $\phi^i$ as axions whenever $N_i \neq 0$.  

In a related matter, let us stress that, in general, the expression \eqref{Knil} \textit{cannot} be used to compute 
the K\"ahler metric, since taking the nilpotent orbit approximation and taking 
derivatives with respect to the moduli does not commute. Nevertheless, 
we can use the full result \eqref{eqn:nilpotentOrbit} for the nilpotent orbit 
of the complete Hodge filtration to compute the K\"ahler metric or derivatives of the 
superpotential. As mentioned above, we will bypass this issue completely
by working directly with the scalar potential \eqref{scalar-potential} and apply 
the approximations to this expression. In the remainder of this section 
we will show how starting from the nilpotent orbit \eqref{eqn:nilpotentOrbit} we can 
derive an approximate scalar potential with an explicit dependence on the axions $\phi^i$. 

\subsection{The boundary $\slt$-structures associated to a degeneration} \label{sec:commSl2}

In the last subsection we have argued that in order to have candidate axion fields $\phi^i$ in complex structure moduli space some of the 
complex structure moduli need to be close to the boundary of this moduli space. Furthermore, we have seen that the  $N_i$ associated to this 
boundary have to be non-vanishing. To study the dynamics of the axions we thus need to evaluate the asymptotic behavior of the scalar potential \eqref{scalar-potential} near the boundary. As will become apparent below the highly non-trivial \emph{$\SLt$-orbit theorem} developed in \cite{MR0382272} and \cite{MR840721} provides the necessary information about the near boundary region to attack this problem. In the following we will briefly review the ingredients of the $\SLt$-orbit theorem needed in this paper.

The main result discussed in section \ref{sec:whyaxions} is the fact that one can associate to each boundary in moduli space 
a nilpotent orbit \eqref{eqn:nilpotentOrbit}. This orbit encodes the Hodge structure \eqref{primHodge-dec} near the boundary 
at $s^i = \infty, i = 1, \ldots, n$. In summary, we have the data $(\Fpol, N_1, \ldots, N_n)$, 
which is regarded as the input of the $\SLt$-orbit theorem. 
The data constructed by the $\SLt$-orbit theorem includes a collection of
\begin{equation} \label{eqn:commsl2FirstAppearance}
  \textrm{commuting } \slt \textrm{-triples:}\qquad (\hN_i, N^0_i, N^+_i)\, ,\quad \textrm{for } i = 1, \ldots, n\, ,
\end{equation}
and a
\begin{equation}
  \textrm{boundary Hodge structure:}\qquad F_\infty\, .
\end{equation}
The commuting $\slt$-triples satisfy the standard relations
\begin{equation}
  [N^0_i, N^{\pm}_i] = \pm 2 N^{\pm}_i\, ,\qquad [N^+_i, N^-_i] = N^0_i\, ,
\end{equation}
and the boundary Hodge structure is again a weight-four pure Hodge structure on the primitive cohomology $\Hp^4(Y, \bbC)$ polarized by the intersection bilinear form $\langle \blank , \blank \rangle$. Unpacking the definition, this means that one is able to define its associated Hodge decomposition
\begin{equation}
  \Hp^4(Y, \bbC) = \bigoplus_{p + q = 4} H_\infty^{p, q}\, ,\quad \textrm{where}\quad H_\infty^{p, q} = F_\infty^p \cap \conj{F}_\infty^q\, ,\quad \textrm{and}\quad H_\infty^{p, q} = \conj{H}_\infty^{q, p}\, ,
\end{equation}
such that the following polarization condition is satisfied
\begin{equation}
  \langle H_\infty^{p, q}, H_\infty^{r, s}\rangle = 0\, ,\quad \textrm{unless } (p, q) = (s, r) \, .
\end{equation}
We can also define its associated Weil operator and Hodge norm according to \cref{eqn:defnWeilOperator,eqn:defnHodgeInnerProduct,eqn:defnHodgeNorm}
\begin{align}
  C_\infty(\alpha)                     &= \im^{p - q}\alpha\, ,\quad \textrm{for } \alpha \in H_\infty^{p, q}\, ,\nonumber\\
  \langle \alpha, \beta \rangle_\infty &= \langle C_\infty \alpha, \conj{\beta} \rangle\, ,\\
  \norm{\alpha}_\infty                 &= \langle \alpha, \alpha \rangle_\infty\, ,\nonumber
\end{align}
where we abbreviate the $F_\infty$ appearing in subscripts as $\infty$ to ease the notational burden.

To proceed further, it is convenient to denote the cumulated sum by braced subscripts. For example, we have
\begin{equation}
  \opY{i} = N^0_1 + \cdots + N^0_i\, .
\end{equation} 
Because the operators $N^0_i$ commute with each other, their cumulated sums $\opY{i}$ also mutually commute $[\opY{i}, \opY{j}] = 0$, thus they have common eigenspaces. In other words, the operators $\opY{i}$ define a multi-grading
\begin{equation} \label{eqn:V-Splitting}
  \Hp^4(Y, \bbR) = \bigoplus_{\Bell = (l_1, \ldots, l_n)} V_{\Bell}\, ,
\end{equation}
where each $V_{\Bell}$ is the simultaneous eigenspace of $\opY{i}$ with eigenvalue $l_i$ \footnote{Note that our convention differs from the convention in \cite{Grimm:2019ixq} by a constant shift of four.}, i.e.
\begin{equation}
  \opY{i}v_{\Bell} = l_i v_{\Bell}\, ,\qquad \textrm{for } v_{\Bell} \in V_{\Bell} \textrm{ and } i = 1, \ldots, n\, .
\end{equation}
The multi-grading \eqref{eqn:V-Splitting} is defined on the real cohomology because the operators $N^0_i$ are elements of $\gR$. In the following, we will also work with complexified multi-grading by allowing complex linear combinations.

The multi-grading \eqref{eqn:V-Splitting} and the boundary Hodge structure $F_\infty$ are compatible with each other in the sense that the multi-grading is orthogonal with respect to the Hodge inner product $\braket{\blank}{\blank}_\infty$, i.e.
\begin{equation}
  \braket{V_{\Bell}}{V_{\Bell'}}_\infty = 0\, ,\qquad \textrm{unless } \Bell = \Bell'\, .
\end{equation}
Furthermore, this Hermitian Hodge inner product becomes a symmetric positive-definite inner product after being restricted to the real cohomology $\Hp^4(Y, \bbR)$, hence we can always choose a real orthonormal basis $\{e_{\Bell}^i\}$ for each $V_{\Bell}$:
\begin{equation} \label{eqn:ONBVl}
  V_{\Bell} = \spanR{e_{\Bell}^i}\, ,\quad \textrm{and}\quad \braket*{e_{\Bell}^i}{e_{\Bell'}^j}_\infty = \delta^{ij}\delta_{\Bell\Bell'}\, .
\end{equation}
Such choice of orthonormal basis will be used in our analysis of the scalar potential \eqref{V-M} near the boundary in section \ref{sec:axionDependentPotential}.

\subsection{Asymptotic form of periods and the scalar potential} \label{sec:normalFormPeriodMapping}

Now we come back to the study of the asymptotic behavior of the scalar potential \eqref{scalar-potential} near the boundary. Our focus will be its first term containing the Hodge metric evaluated at $F(t)$. We will first introduce a normal form of the period mapping $F(t)$, which factors $F(t)$ into nice pieces. Moreover, each factor in the normal form has a good limiting property near the boundary where $s^i = \infty$. These limits combine into each other, yielding an asymptotic form of the scalar potential that will be the object studied in section \ref{sec:axionDependentPotential}.

In order to write down the normal form, we need again the data of the nilpotent orbit $(\Fpol, N_1, \ldots, N_n)$ defined in \eqref{eqn:defnLogMonodromy} and \eqref{eqn:nilpotentOrbit}. In addition, we require an extra piece of information: a holomorphic function $\Gamma(z)$ valued in the Lie algebra $\fg_\bbR$. This function satisfies certain properties that are reviewed in appendix \ref{app:commutingsl2}. For the moment we only need to know its existence. Then the \emph{normal form of the period map} \cite{MR1042802} is given by
\begin{equation} \label{eqn:periodMapNormal}
  F(t) = e^{t^i N_i} e^{\Gamma(z)} F_0 = e^{\phi^i N_i} e^{\im s^i N_i} e^{\Gamma(z)} F_0\, ,
\end{equation}
where we remind the reader that the $t^j$-dependence in the $\Gamma(z)$-function is introduced via $z^j = e^{2\pi\im t^j}$ as in \eqref{eqn:defnOfti}.
The normal form provides a convenient factorization of the period mapping into group elements $e^{\phi^i N_i} \in G_\bbR$ and $e^{\im s^i N_i}$,$e^{\Gamma(z)}\in G_\bbC$ acting on the filtration $F_0$. 

We will now introduce a natural action of $G_\bbC$ on the Hodge norm, which will not only let us factor out the axion dependence in the scalar potential, but also turn the general study of Hodge norm in the bulk near the boundary into the study of the Hodge norm induced by $F_\infty$ at the boundary. This action is defined as follows.
For every element $g \in G_\bbC$ and form $\alpha, \beta \in \Hp^4(Y, \bbC)$, there is a tautological relation
\begin{equation}
  \braket{\alpha}{\beta}_{F} = \braket{g \alpha}{g \beta}_{g F}\, .
\end{equation}
This relation then equips the following action of $g$ on the Hodge norm of $\alpha$
\begin{equation}
  \Vert \alpha \Vert_{g F} = \Vert g^{-1} \alpha \Vert_{F}\, .
\end{equation}
These properties of the Hodge norm and the normal form of the period mapping allow us 
to factor out the axion dependence in the scalar potential. More precisely, using the action of $G_\bbC$ on the Hodge norm we 
can write 
\begin{equation} \label{split-axion-off}
  \norm{G_4}_{F(t)} = \norm{e^{-\phi^i N_i} G_4}_{e^{\im s^i N_i} e^{\Gamma(z)} F_0}\, .
\end{equation}
In accordance with the nilpotent orbit theorem discussed in section \ref{sec:whyaxions} (see appendix \ref{app:commutingsl2} for more details) 
near the boundary $s^i = \infty$, the term $e^{\Gamma(z)}$ will provide exponentially suppressed corrections. It is in this sense that near the boundary the dependence of $\norm{G_4}_{F(t)}$ in the axion $\phi^i$ and saxion $s^i$ are separated. Note also that the operator $e^{-\phi^i N_i}$ being an element in the group $G_\bbR$ by definition preserves the polarization form: $\langle e^{-\phi^i N_i} G_4,  e^{-\phi^i N_i} G_4 \rangle = \langle G_4, G_4 \rangle$. These facts instruct us to define the modified $G_4$-flux including the axions in \cite{Grimm:2019ixq}
\begin{equation} \label{def-rho}
  \rho(\phi, G_4) = e^{-\phi^i N_i} G_4\, .
\end{equation}
This redefinition has also been motivated and discussed intensively in \cite{Bielleman:2015ina,Herraez:2018vae,Marchesano:2019hfb}. In this paper we find it more convenient to not use this redefinition $\rho(\phi, G_4)$, but rather always directly display $G_4$. In summary, the scalar potential expressed using the split \eqref{split-axion-off} takes the form 
\begin{equation} \label{eqn:potentialInRho}
  V(t) = \frac{1}{\cV_4^3}\bigg[\norm{e^{-\phi^i N_i} G_4}_{e^{\im s^i N_i} e^{\Gamma(z)} F_0}^2 - \langle G_4, G_4 \rangle\bigg]\, .
\end{equation}

Let us now turn to a more in-depth study of the asymptotic form of the scalar potential \eqref{eqn:potentialInRho}
near the boundary. In higher-dimensional moduli spaces, the asymptotic behavior can depend on the path along which one approaches the boundary $s^i = \infty$. In order to regulate this, we need to introduce the \emph{growth sector}
\begin{equation} \label{eqn:defnGrowthSector}
  \cR_{1\cdots n} = \left\{ \frac{s^1}{s^2} \geq \lambda, \ldots, \frac{s^{n - 1}}{s^n} \geq \lambda, s^n > \lambda \right\}\, ,
\end{equation}
where we consider $\lambda \geq 1$. The definition of the growth sector binds with an ordering of the variables $t^i$. Setting $\lambda =1$ 
one can cover the entire neighborhood of the boundary by growth sectors obtained by considering every ordering in 
the variables $t^i$. In the following we will focus on one of these sectors, namely $\cR_{1\cdots n}$, after possibly renaming the 
coordinates. Gluing these sectors together can be a non-trivial task, but is not of importance in the remainder of this work. 
Note that eventually we will work in the `strict asymptotic regime', i.e.~we will assume $\lambda \gg 1$. 

In order to analyze the asymptotic behavior of the Hodge norm, i.e., the first $t$-dependent term in \eqref{eqn:potentialInRho}, we will introduce an operator $e(s)$ following \cite{MR1042802}. It is defined by
\begin{equation} \label{eqn:defnEOperatorInS}
  e( s^1, \ldots, s^n ) = \exp\left\{ \hf ( \log s^r ) N^0_r \right\}\, ,
\end{equation}
where $r$ is being summed from $1$ to $n$. One of the motivations behind the introduction of the $e(s)$-operator is the property
\begin{equation} \label{eqn:exe-1}
  e(s) e^{ \phi^j N_j } e(s)^{-1} = \exp\Bigg\{ \sum_{j = 1}^n \frac{\phi^j}{s^j} \Bigg[ \hN_j + \sum_{\ua^j > 0} \frac{N'_{j, \ua^j}}{\big(\frac{s^1}{s^2}\big)^{\alpha^j_{1}/2} \cdots \big(\frac{s^{j - 1}}{s^j}\big)^{\alpha^j_{j - 1}/2}} \Bigg] \Bigg\}\, ,
\end{equation}
where $N_j$ is the log-monodromy operator introduced in \eqref{eqn:defnLogMonodromy}, $\hN_j$ is the corresponding lowering operator in the commuting $\slt$-triples introduced in \eqref{eqn:commsl2FirstAppearance}. The operators $N'_{j, \ua^j}$ are nilpotent operators living in $\gR$. The derivation of this expression together with the precise definition of the operators $N'_{j, \ua^j}$ can be found in appendix \ref{app:commutingsl2}. The main point is that if one moves towards the boundary $s^i = \infty$ within the growth sector $\cR_{1\cdots n}$, while not keeping the candidate axion $\phi^i$ finite, the terms proportional to $N'_{j, \ua^j}$ are polynomially suppressed. This implies that in the limit of large $s^i$ we find 
within $\cR_{1\cdots n}$ the limiting behavior  
\begin{align} \label{eqn:exe-1limit}
  e(s)e^{ \phi^j N_j }e(s)^{-1} &\to \exp\bigg( \sum_{j = 1}^n \frac{\phi^j}{s^j} \hN_j \bigg)\, , \\ 
  e(s)e^{ \im s^j N_j }e(s)^{-1} &\to \exp\bigg(\im \sum_{j = 1}^{n} \hN_j\bigg)\, , \label{eqn:eye-1limit}
\end{align}
where the second line can be obtained from the first by formally setting $\phi^j = \im s^j$. 
If we look at the normal form of the period mapping \eqref{eqn:periodMapNormal}, we see that some limiting expressions related to $e^{\Gamma(z)}$ and $F_0$ are also needed. We state the correct form of the limits here, and refer to appendix \ref{app:commutingsl2} for their derivation
\begin{align}
  e(s) e^{\Gamma(z)} e(s)^{-1}                           & \to 1\, ,\nonumber\\
  \exp\bigg(\im \sum_{j = 1}^{n} \hN_j\bigg) e(s) F_0 & \to F_\infty\, , \label{eqn:limitingPropertiesWithAde}
\end{align}
where $F_\infty$ is the boundary Hodge filtration.

Combining the normal form of the period mapping \eqref{eqn:periodMapNormal} and \cref{eqn:exe-1limit,eqn:eye-1limit,eqn:limitingPropertiesWithAde} gives us the following limiting expression of the period mapping $F(t)$ near the boundary $s^i = \infty$ in the growth sector $\cR_{1\cdots n}$
\begin{equation} \label{eqn:periodMapAsymptotic}
  F(t) \to e(s)^{-1}\exp\bigg( \sum_{j = 1}^n \frac{\phi^j}{s^j} \hN_j \bigg) F_\infty\, .
\end{equation}

The above result can be applied immediately to the study of the scalar potential \eqref{eqn:potentialInRho}. Combining the action of $G_\bbC$ on the Hodge norm and the asymptotic expression of the period mapping \eqref{eqn:periodMapAsymptotic}, we have \footnote{As usual, the symbol $\sim$ indicates that the quantities on both sides approximate each other increasingly well as one move towards the considered boundary.}
\begin{align}
  \norm{e^{-\phi^i N_i} G_4}_{e^{\im s^i N_i} e^{\Gamma(z)} F_0} & = \norm{e(s) e^{-\phi^i N_i} G_4}_{e(s) e^{\im s^i N_i} e^{\Gamma(z)} F_0}
                                                                 \sim \bigg\| \exp\bigg( \sum_{j = 1}^n -\frac{\phi^j}{s^j} \hN_j \bigg) e(s)\, G_4\bigg\| _\infty\, .\label{eqn:equivalenceOldNew}
\end{align}
So in the strict asymptotic regime $\lambda \gg 1$ in \eqref{eqn:defnGrowthSector}, the scalar potential \eqref{eqn:potentialInRho} has the following asymptotic form
\begin{equation} \label{eqn:asymScalarPotential}
  \boxed{ \rule[-.6cm]{.0cm}{1.4cm}  \quad V(t) \sim \frac{1}{\cV_4^3}\Bigg[\bigg\| {\exp\bigg(\sum_{j = 1}^n -\frac{\phi^j}{s^j} \hN_j\bigg)e(s) G_4} \bigg\| _\infty^2 - \langle G_4, G_4 \rangle\Bigg]\, . \quad }
\end{equation}
This asymptotic form of the scalar potential is the main result from asymptotic Hodge theory that we will use in the following. The important thing 
to note is that all dependence in \eqref{eqn:asymScalarPotential} on the axions $\phi^i$ and the saxions $s^i$ is explicit and given in terms of the boundary structure. In particular, the norm $\| \cdot \|_\infty$ is both well-defined and independent of $t^i$, but can depend still on the 
coordinates not considered to be near the boundary. Combined with the underlying 
$\slt$-structure we can use this form of the potential to tackle the stabilization of saxions in the presence of large displacement of axions.

Let us close this section with a short comment on the apparent discrepancy between \eqref{eqn:asymScalarPotential} and our result in section 7 of \cite{Grimm:2019ixq}. The reason of the discrepancy is that in \cite{Grimm:2019ixq} we replaced all $N_i$ in the expression of $\rho(\phi, G_4)$, given in \eqref{def-rho}, by their commuting $\slt$-counterparts $\hN_i$. This was done to simplify the computation, but it neglected the contributions from the difference between $N_i$ and $\hN_i$. Had the replacement not been done, the two approaches are equivalent because
\begin{equation}
  \norm{ e(s) \rho(\phi, G_4) }_\infty \sim \bigg\| \exp\bigg( \sum_{i = 1}^n -\frac{\phi^i}{s^i} \hN_i \bigg) e(s) \, G_4 \bigg\|_\infty\, ,
\end{equation}
which is the step deriving \eqref{eqn:equivalenceOldNew}. The left-hand-side of the above equation is the object studied in \cite{Grimm:2019ixq} whereas the right-hand-side is the object studied in this note.

\section{Axion backreaction on the saxion vacuum} \label{sec:axionDependentPotential}

In this section we study the backreaction of a large displacement of 
an axion, denoted by $\phi^k$, on the saxion vacuum expectation values. This forces us to 
discuss moduli stabilization within the general 
scalar potential \eqref{eqn:asymScalarPotential}, which is a very hard problem. In particular, 
while we know the field dependence of \eqref{eqn:asymScalarPotential} on the fields $t^i$, we have no control 
over its dependence on the coordinates of complex structure moduli space not considered close to the boundary. 
Here the power of asymptotic Hodge theory comes to the rescue, since it allows us to control at least the positivity 
properties and the hierarchy of certain couplings that cannot be further specified. 
We will then focus on one pair of axion and saxion $(\phi^k, s^k)$ and study the stabilization of 
the saxion $s^k$ via the potential \eqref{eqn:asymScalarPotential} to its vacuum expectation value 
$\svev{k}$. To simplify notation we use in this section the definition:
\beq
    s \equiv s^k\ , \qquad \phi \equiv \phi^k\ . 
\eeq
We then find that whenever we try to fix $s$ and assume that the associated axion $\phi$ 
is large, we find the following universal relation
\begin{equation} \label{eqn:wantedBackreaction}
 \boxed{ \rule[-.4cm]{.0cm}{1cm} \quad  \svev{}(\phi) \sim c \, \phi^\gamma + \cO\bigg( \frac{1}{\phi} \bigg)\, , \quad }
\end{equation}
where $c$ is a positive number, and $0 < \gamma \le 2$ is a rational number. While our expression is slightly more general, the relation \eqref{eqn:wantedBackreaction} 
shows that one  always encounters the type of backreaction that was found in \cite{Baume:2016psm,Valenzuela:2016yny,Blumenhagen:2017cxt}. In fact, we will argue in the end of section \ref{saxion_vacuum} that in most cases one indeed has $\gamma = 1$, and in general $\gamma < 1$ can appear if some special choices of parameters are allowed. The $\gamma > 1$ cases are even rarer in the sense that only two valid special cases with $\gamma = 2$ are found. In other words, the fact that one cannot 
displace the axion by very large values without destabilizing the saxion is a consequence of the boundary $\slt$-structure introduced in the last section.
We stress, however, that we will not be able to make statements about the precise value of $c$ and its dependence on the fluxes 
and other moduli. While $c$ cannot be made zero, we will not exclude the possibility that it can be made small by fine-tuning leading to a somewhat delayed backreaction \cite{Blumenhagen:2014nba,Hebecker:2014kva,Valenzuela:2016yny,Landete:2017amp}. A short discussion on the dependency of $c$ and the flux numbers can also be found at the end of section \ref{saxion_vacuum}.

Our study relies heavily on the action of the commuting $\slt$-triples on the cohomology $\Hp^4(Y, \bbR)$, hence we will start with reviewing some elementary facts about $\slt$-representations in section \ref{repofsl(2)}. The boundary $\slt$-structure then allows us to bring the 
asymptotic scalar potential into a convenient form and can be extremized in the limit of large axion.
This amounts to solving a one-parameter family of one-variable polynomial equations and study how the root depends on the parameter in certain limits. This kind of problem is exactly studied by a well-known mathematical tool called the Puiseux expansion, whose information is registered in a pictorial way in the so-called Newton diagram. We will review the notion of Puiseux expansions and their associated Newton diagrams in section \ref{sec:NewtonPuiseux}.
Having introduced these additional tools we turn in section \ref{saxion_vacuum} to the detailed study of the asymptotic scalar potential and its extremization condition. With the help of the boundary $\slt$-structure and the Newton diagram, we will show that almost all flux configurations will give an axion backreaction behavior of the form \eqref{eqn:wantedBackreaction}. We will also enumerate all possible flux configurations that can potentially generate a backreaction behavior different from \eqref{eqn:wantedBackreaction}. These flux configurations are then studied case-by-case 
in section \ref{app:badCases}. Interestingly, none of them actually yields a valid solution, in the sense that each case generates a saxion vev leading term that is either negative or imaginary. We conclude this section with a list of bad cases, where some explicit examples are also provided.

Before we start, let us state again the scalar potential \eqref{eqn:asymScalarPotential} focusing only on a single axion-saxion pair $(\phi, s)$, but allowing 
for a slight generalization with an overall factor. More precisely, we will consider in the following the scalar potential 
\begin{equation} \label{eqn:kScalarPotential}
  \boxed{ \rule[-.6cm]{.0cm}{1.4cm} \quad V(\phi, s) := \frac{1}{s^\alpha}\Bigg[\norm{\exp\bigg(-\frac{\phi}{s} \hN \bigg) e(s) G_4}_\infty^2 - A_{\textrm{loc}}\Bigg]\, , \quad }
\end{equation}
where we have set $ N^- \equiv \hN_k$ and recall that the $e(s)$-operator is defined in \eqref{eqn:defnEOperatorInS}. The localized contribution that does not depend on the complex structure moduli is collectively denoted by $A_{\textrm{loc}}$. Following~\cite{Grimm:2019ixq}, we have included the overall scaling $1/s^\alpha$ with an undetermined power $\alpha$. This factor can be thought of as arising from $\cV_4$ in \eqref{eqn:asymScalarPotential} and enables a comparison between the F-theory potential and the IIA scalar potential \cite{Grimm:2004ua,DeWolfe:2005uu}. Around the weak coupling limit of  F-theory, the value of $\alpha$ is known to be $3$ when $s$ is related to the Type IIA dilaton as discussed in detail in \cite{Grimm:2019ixq}. In general 
it is not known how large $\alpha$ is near other singularities. However, requiring that the scalar potential \eqref{eqn:kScalarPotential} is finite as $s \to \infty$ restricts the possible range of $\alpha$. We will find that after imposing this restriction the backreaction \eqref{eqn:wantedBackreaction} is universal.

\subsection{A brief review of representations of the $\slt$-algebra} \label{repofsl(2)}

It turns out that asymptotics of \eqref{eqn:kScalarPotential} with respect to $(\phi, s) \equiv (\phi^k, s^k)$ only depends on the behavior of the $G_4$-flux under the action of the $k$-th commuting $\slt$-triple, whose lowering and number operators are denoted by $(\hN, N^0)$, respectively. We will abuse the notation and denote the $k$-th $\slt$-triple $(\hN_k, N_k^0, N_k^+)$ just by $(\hN, N^0)$. In this subsection let us recall some elementary facts of $\slt$-algebra representations \cite{MR1808366}. We work over the real numbers since the $\slt$-triple and $G_4$-flux are real, nevertheless the theory holds for complex representations as well. We align with the notation in section 4.2 of \cite{Grimm:2020cda}.

For any integer $d \ge 0$, there is a $(d + 1)$-dimensional irreducible representation $\cW^d$ of the $\slt$-algebra $(\hN, N^0)$. One can specify a special state $\ket{d, d}$ in $\cW^d$ called the highest weight state of weight $d$. It satisfies the property 
\begin{equation}
  (\hN)^d\ket{d, d} \ne 0\quad \textrm{and}\quad (\hN)^{d + 1}\ket{d, d} = 0\, .
\end{equation}
A basis of the representation $\cW^d$ can then be constructed out of the highest weight state $\ket{d, d}$ and the lowering operator $\hN$ as follows
\begin{equation} \label{eqn:basisOfWd}
  \cW^d = \spanR{\ket{d, d}, \ket{d, d - 2}, \ldots, \ket{d, -d}}\, ,
\end{equation}
where
\begin{equation}
  \ket{d, d - 2n} = \frac{1}{n!}(\hN)^n \ket{d, d}\, ,
\end{equation}
for $n = 0, \ldots, d$. These vectors are also eigenstates of the $N^0$ operator, satisfying
\begin{equation}
  N^0 \ket{d, l} = l \ket{d, l}\, .
\end{equation}
We call the eigenvalue $l$ the weight of the state $\ket{d, l}$. This state also satisfy
\begin{equation}
  (\hN)^{\frac{d + l}{2}}\ket{d, l} \ne 0\quad \textrm{and}\quad (\hN)^{\frac{d + l}{2} + 1}\ket{d, l} = 0\, .
\end{equation}
Note that by construction, $d + l$ is always an even non-negative number.

Now we make contact with the boundary $\slt$-structure. In equation \eqref{eqn:commsl2FirstAppearance}, a series of commuting $\slt$-triples is introduced at the boundary of the complex structure moduli space. These commuting $\slt$-algebras act on $\Hp^4(Y, \bbR)$ and, in particular, turn $\Hp^4(Y, \bbR)$ into a real representation of the $k$-th $\slt$-algebra $(\hN, N^0)$. According to the above discussion of representations of $\slt$-algebras, $\Hp^4(Y, \bbR)$ enjoys the following decomposition
\begin{equation} \label{eqn:isotypicalV}
  \Hp^4(Y, \bbR) = \bigoplus_{d = 0}^{4} \cW[d]\, ,
\end{equation}
where
\begin{equation}
  \cW[d] = \cW^d_{1} \oplus \cdots \oplus \cW^d_{\mu_d}
\end{equation}
consists of $\mu_d$ copies of irreducible representations $\cW^d_{i_d}$ of dimension $d + 1$. Different highest weight states with the same $d$-label are distinguished by the index $i_d = 1, \ldots, \mu_d$: they are denoted by $\ket{d, d; i_d} \in \cW^d_{i_d}$ and their descendants are denoted similarly by $\ket{d, l_d; i_d}$.
One also sees that each basis vector in \eqref{eqn:basisOfWd} has a well-defined eigenvalue under the action of $N^0$. Hence we relate the orthonormal basis \eqref{eqn:ONBVl} adapted to the multi-grading to the basis vectors in \eqref{eqn:basisOfWd} in a one-to-one manner. We fix the basis \eqref{eqn:basisOfWd} in this way, so that two basis vectors in \eqref{eqn:basisOfWd} are orthogonal to each other unless they carry identical indices $d, l_d$ and $i_d$.

\subsection{A brief review of the Puiseux expansion} \label{sec:NewtonPuiseux}

To determine the backreacted saxion vacuum expectation value, we need to study how the root of a one-parameter family of polynomial equations change with respect to the parameter. This type of question can be studied expanding the solution into Puiseux series. In this subsection we briefly review the use of the Puiseux expansion and Newton diagram. We will not show any proof of the facts and the interested reader can find the proof in \cite{MR0033083}.

For simplicity, we work over the complex numbers in this subsection so that every polynomial always has roots. When we apply the Puiseux expansion to analyze the axion backreaction, we will always require the existence of a vacuum. This means that the polynomial arising from the first derivatives of the scalar potential, see \eqref{eqn:dVkInKSplitting} below, is assumed to have a real root. The method of expanding the root into a Puiseux series also applies in such circumstances.

The Puiseux expansion studies generalized polynomial equations in two variables $F(s, \hphi) = 0$, with the variable $\hphi$ being distinguished in the sense that the powers of $\hphi$ are allowed to be negative. The equation $F(s, \hphi) = 0$ can also be regarded as a one-variable polynomial equation with a parameter $\hphi$. Around $\hphi = 0$, any $s$ satisfying $F(s, \hphi) = 0$ can be regarded as a function of $\hphi$. The series representation of $s(\hphi)$ is given by the Puiseux expansion, which is a fractional power series. More precisely, let us assume the following form of a generalized polynomial
\begin{equation} \label{eqn:generalPolynomialEquation}
  F(s, \hphi) = a_0(\hphi) + a_1(\hphi)s + \cdots + a_n(\hphi)s^n\, ,
\end{equation}
where every $a_i(\hphi)$ is a polynomial in $\hphi$ and $1/\hphi$ with complex coefficients
\begin{equation}
  a_i(\hphi) = \sum_j a_{ij} \hphi^j\, .
\end{equation}
Then the Puiseux expansion states that, near $\hphi = 0$, any root of the equation $F(s, \hphi) = 0$, regarded as a function $s = s(\hphi)$, can be expanded as
\begin{equation} \label{eqn:generalNewtonPuiseux}
  s(\hphi) = \frac{c}{\hphi^\gamma} + \sum_{i = 0}^\infty c_i \hphi^{i/m}\, ,
\end{equation}
where $c$ and $c_i$ are complex numbers, $\gamma$ is a rational number and $m$ is a positive integer.
Our focus is on the leading power $\gamma$. Knowing that any $s$-root must admit a fractional power series expansion \eqref{eqn:generalNewtonPuiseux}, the determination of $\gamma$ is standard: One inserts the expansion \eqref{eqn:generalNewtonPuiseux} back to the original equation \eqref{eqn:generalPolynomialEquation} and solves for the $\gamma$ that makes the lowest order terms cancel. This whole procedure was encoded by Newton into an intuitive gadget called the Newton polygon which we introduce next. We would like to comment that despite determining $\gamma$ is straightforward once one has the Ansatz \eqref{eqn:generalNewtonPuiseux}, the significance of the Puiseux expansion lies in the proof of convergence of the fractional series \cite{MR0033083}.

\begin{figure}[H]
  \centering
  \begin{tikzpicture}[scale=1]
    \draw[->] (0, -2.5) -- (0, 0.5) node[above] {$j$};
    \draw[->] (0, 0) -- (5.5, 0) node[right] {$i$};
  
    \draw[double] (3, -2) -- (5, 0);
    \draw[double] (0, -2) -- (3, -2);

    \draw[fill] (0, -2) circle[radius=0.05] node[left] {$-2$};
    \draw[fill] (2, 0) circle[radius=0.05] node[above] {$2$};
    \draw[fill] (3, -2) circle[radius=0.05];
    \draw[fill] (5, 0) circle[radius=0.05] node[above] {$5$};
  \end{tikzpicture}
  \caption{The Newton diagram $\Delta(F)$ of the polynomial \eqref{eqn:examplePolynomial} consists of the four solid dots shown in this figure. The corresponding Newton polygon, the lower convex hull of the Newton diagram, is labelled by double lines. The vertical axis labels the powers of $\hat \phi$ while the horizontal axis labels the powers of $s$.} \label{fig:exampleNewton}
\end{figure}
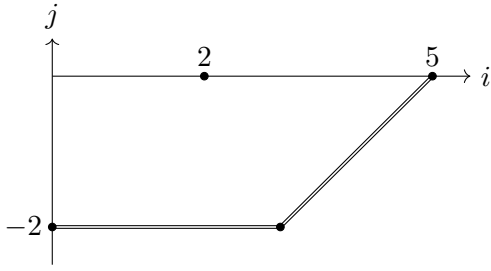

To determine $\gamma$ pictorially, we need to first define the \emph{Newton diagram} $\Delta(F)$ of $F(s, \hphi)$ as follows
\begin{equation}
  \Delta(F) = \left\{ (i, j) \in \bbR^2 \mid a_{ij} \ne 0 \textrm{ in } F(s, \hphi) \right\}\, ,
\end{equation}
which simply consists of the dots $(i, j)$ on the plane such that a term $s^i \hphi^j$ with non-vanishing coefficient $a_{ij} \ne 0$ appears in the generalized polynomial $F(s, \hphi)$. By our assumption on the polynomial $F$, its Newton diagram $\Delta(F)$ will only occupy the half plane $i \ge 0$. Then the \emph{Newton polygon} of $F$ is defined to be the lower convex hull of $\Delta(F)$.
To illustrate the definition with an example let us consider the  polynomial
\begin{equation} \label{eqn:examplePolynomial}
  F(s, \hphi) = \frac{a_1}{\hphi^2} + a_2 s^2 + \frac{a_3}{\hphi^2} s^3 + a_4 s^5\, ,
\end{equation}
where $a_1, a_2, a_3$ and $a_4$ are non-zero numbers. This polynomial arises in a specific $2$-moduli degeneration of Calabi-Yau fourfolds.\footnote{The degeneration is classified as of type $\rI_{01} \to \rV_{22}$ with the $G_4$ flux chosen to be $G_4 = g_{42} v_{42} + g_{34} v_{34}$ in the convention of \cite{Grimm:2019ixq}.} The Newton diagram and Newton polygon is shown in Figure \ref{fig:exampleNewton}.

The Newton polygon consists of several segments. Each segment with slope $\gamma$ determines a possible leading exponent $\gamma$ in the Puiseux expansion \eqref{eqn:generalNewtonPuiseux}. In our example \eqref{eqn:examplePolynomial}, with Newton polygon shown in Figure \ref{fig:exampleNewton}, the polygon consists of two segments with slopes $0$ and $+1$. So around $\hphi = 0$, the equation has the following roots represented by Puiseux expansions
\begin{align}
  s_1(\hphi) & = \frac{c_1}{\hphi} + \sum_{i = 0}^\infty c_{1, i} \hphi^{i/m}\, ,\\
  s_2(\hphi) & = c_2 + \sum_{i = 1}^\infty c_{2, i} \hphi^{i/n}\, ,
\end{align}
where $m, n$ are positive integers. In our application, the $\hphi$ here actually stands for the inverse of an axion field $1/\phi$ and $s$ stands for the corresponding saxion partner $s$. Hence the first solution $s_1(\hphi)$ represents a linear-backreacted saxion vacuum expectation value, while the second solution $s_2(\hphi)$ stays finite at large $\phi$.

\subsection{Determining the backreacted saxion vacuum} \label{saxion_vacuum}
We now have all the tools we need to attack the axion backreaction problem. Using the $\slt$-basis, we will first expand the asymptotic scalar potential \eqref{eqn:kScalarPotential} into a generalized polynomial in $(s,\phi)$. Then we will look closer at the shape of the Newton diagram to deduce the leading term in the backreacted saxion vev $\svev{}(\phi)$.

Let us begin by decomposing the $G_4$-flux according to \eqref{eqn:isotypicalV} as
\begin{equation}
  G_4 = \sum_{d = 0}^4 \sum_{i_d = 1}^{\mu_d} \sum_{n_d = 0}^{d} g_{d, d - 2n_d; i_d} \ket{d, d - 2n_d; i_d}\, ,
\end{equation}
where $g_{d, d - 2n_d; i_d}$ is the flux-component of the highest weight representation $\cW^d$ with weight $d - 2n_d$. Note that the $e(s)$-operator defined in \eqref{eqn:defnEOperatorInS} acts on a basis state $\ket{d, l_d; i_d}$ by scalar multiplication
\begin{equation} \label{eqn:eOpKOnly}
  e(s) \ket{d, l_d; i_d} = s^{\frac{l_d}{2}} \hat{f}_{d, l_d; i_d} \ket{d, l_d; i_d}\, ,
\end{equation}
where $\hat{f}_{d, l_d; i_d}$ can be a non-vanishing function\footnote{Note that in our multi-grading notation in \cite{Grimm:2019ixq}, the $l_d$ here corresponds to the $l_k - l_{k - 1}$ index of $\ket{d, l_d; i_d}$ in the \cite{Grimm:2019ixq}.} of all other complex structure coordinates.
Then by a direct computation, the asymptotic scalar potential is found to be
\begin{align}
  V(\phi, s) & = \frac{1}{s^\alpha}\left[\norm{\exp\bigg(-\frac{\phi}{s} \hN\bigg) e(s) G_4}_\infty^2 - A_{\textrm{loc}}\right]\\
             & = \frac{1}{s^\alpha}\left[\sum_{d = 0}^4 \sum_{i_d = 1}^{\mu_d} \sum_{n_d = 0}^{d} \sum_{b_d = n_d}^{d} \binom{b_d}{n_d}^2 s^{d - 2b_d} \phi^{2(b_d - n_d)} g^2_{d, d - 2n_d; i_d} \hat{f}^2_{d, d - 2n_d; i_d} - A_{\textrm{loc}}\right]\nonumber\, ,
\end{align}
where in the last step we have used the orthonormal property of the basis states. We have highlighted the dependency of $V$ only on the pair $(\phi, s)$ and moved the dependencies on the other complex structure moduli into the various $\hat{f}$. Note that the flux number $g_{d, d - 2n_d; i_d}$ is always accompanied by the non-vanishing function $\hat{f}_{d, d - 2n_d; i_d}$. We therefore introduce the redefinition $\hg_{d, d - 2n_d; i_d} = g_{d, d - 2n_d; i_d} \hat{f}_{d, d - 2n_d; i_d}$ to shorten the equations. The expanded asymptotic scalar potential now takes the form
\begin{equation}
  V(\phi, s) = \frac{1}{s^\alpha}\left[\sum_{d = 0}^4 \sum_{i_d = 1}^{\mu_d} \sum_{n_d = 0}^{d} \sum_{b_d = n_d}^{d} \binom{b_d}{n_d}^2 s^{d - 2b_d} \phi^{2(b_d - n_d)} \hg^2_{d, d - 2n_d; i_d} - A_{\textrm{loc}}\right]\, .\label{eqn:VkInKSplitting}
\end{equation}
With the expansion of the asymptotic scalar potential, we can impose a constraint on the undetermined exponent $\alpha$. We require that the potential $V(\phi, s)$ does not blow up in the limit $s \to \infty$. Physically this means that the boundary of the moduli space is viable, not obstructed by the potential. This translates to the constraint that
\begin{equation} \label{eqn:constraintOnAlpha}
  \alpha \ge \max_{g_{d, d - 2n_d; i_d} \ne 0}\{ d - 2n_{d}, 0\}\, ,
\end{equation}
which says that $\alpha$ should not be smaller than the largest weight carrying a non-zero flux component appearing in $G_4$.

Since we are going to study the situation where $\phi$ is large, it is instructive to change variable to $\hphi = 1/\phi$ so that the limit $\phi \to \infty$ corresponds to $\hphi \to 0$. In this coordinate $(s, \hphi)$, we denote the derivative of $V$ with respect to $s$ by $F(s, \hphi)$. A simple computation leads to
\begin{equation} \label{eqn:dVkInKSplitting}
  F(s, \hphi) = \frac{\partial V}{\partial s} = \frac{1}{s^\alpha}\left[\frac{\alpha A_{\rm loc}}{s} + \sum_{d = 0}^4 \sum_{n_d = 0}^{d}  F_{d, n_d}(s, \hphi)\right]\, ,
\end{equation}
where we have grouped the summand according to $d$ and $n_{d}$ as
\begin{equation} \label{eqn:dVkInKSplittingComponent}
  F_{d, n_d}(s, \hphi) = \sum_{b_d = n_d}^{d} \cC_{d, n_d; b_d} s^{d - 2b_d - 1} \hphi^{-2(b_d - n_d)}\, ,
\end{equation}
and the coefficient is given by
\begin{equation} \label{eqn:coefficientC}
  \cC_{d, n_d; b_d} = \sum_{i_d = 1}^{\mu_d} (-\alpha + d - 2b_d) \binom{b_d}{n_d}^2 \hg^2_{d, d - 2n_d; i_d}\, .
\end{equation}
Note that in general $\cC_{d, n_d; b_d} \ne 0$ unless for special combinations of $\alpha, d$, and $b_d$. For the sake of discovering the general properties of the solution that are directly related to the boundary $\slt$-structure, let us temporarily assume that the value of $\alpha$ does not make any of the coefficients $\cC_{d, n_d; b_d}$ vanish. We will discuss the consequence of special choices of $\alpha$ that kill some coefficients $\cC_{d, n_d; b_d}$ in the end of this subsection.

Now we have the stage set up: In order to study the backreacted saxion vev, we should solve the extremization condition $F = 0$ with the polynomial $F$ given in \eqref{eqn:dVkInKSplitting} by the Puiseux series. In order to find the Puiseux series, we should draw the Newton diagram of the polynomial $F$. It turns out that it is more instructive to first look at the Newton diagram of every $F_{d, n_d}$ and then assemble them together into the Newton diagram of $F$.

For every $d$ and $n_d$, we first note that not every power of $s$ in $F_{d, n_d}$ is positive. In order to apply the method of Puiseux expansions, we need to pull out sufficient power of $1/s$ so that the remaining polynomial has only positive powers on $s$. So we define
\begin{equation}
  \tilde{F}_{d, n_d} = s^{d + 1} F_{d, n_d} = \sum_{b_d = n_d}^d \cC_{d, n_d; b_d} s^{2(d - b_d)} \hphi^{-2(b_d - n_d)}\, . \label{eqn:tildeF_d_nd}
\end{equation}
Let us check the shape of $\Delta(\tilde{F}_{d, n_d})$. Denote the powers of $s$ by $a$, the powers of $\hphi$ by $b$, and draw the Newton diagram on the $(a, b)$ plane. It is easy to see that it has all dots aligned along the line segment of slope $+1$ intersecting the $a$-axis at $(2(d - n_d), 0)$ and the $b$-axis at $(0, -2(d - n_d))$. This immediately prompts the following important observation as we fix a $d$ and perform the sum over $n_d$.

Let us fix a $d$, and let $\tilde{n}_d$ be the smallest $n_d$ such that $g_{d, d - 2n_d; i_d} \ne 0$. If there are multiple possible $i_d$ for such flux components, we just arbitrarily pick one since only the values of $d$ and $n_d$ matter. This flux component corresponds to one of the highest weight components $g_{d, d - 2\tilde{n}_d; i_d}$ of $G_4$ inside $\cW[d]$. The sum over $n_d$ with the fixed $d$ takes the following form
\begin{equation} \label{eqn:sumOverND}
  \sum_{n_d = \tilde{n}_d}^d F_{d, n_d} = \frac{1}{s^{d + 1}} \sum_{n_d = \tilde{n}_d}^d \tilde{F}_{d, n_d}\, .
\end{equation}
The Newton diagram of each $\tilde{F}_{d, n_d}$ has been analyzed above. We notice that if we take the lower convex hull to find the Newton polygon of \eqref{eqn:sumOverND} at this stage, the polygon only depends on the lowest line in the Newton diagram, i.e. the one generated by $\tilde{F}_{d, \tilde{n}_d}$. This implies that when we further sum over $d$ as in \eqref{eqn:dVkInKSplitting} and aiming to find the Newton polygon of the entire $F$, only the Newton diagram of $\tilde{F}_{d, \tilde{n}_d}$ matters for every $d$. This observation instructs us to just focus on the highest weight component $g_{d, d - 2\tilde{n}_d; i_d}$ for every $d$.

With the above discussion, the derivative of $V$ \eqref{eqn:dVkInKSplitting} becomes
\begin{equation} \label{eqn:sum_of_tildeF_d_tildend}
  F(s, \hphi) = \frac{\alpha A_{\rm loc}}{s^{\alpha + 1}} + \sum_{d = 0}^4 \frac{\tilde{F}_{d, \tilde{n}_d}(s, \hphi)}{s^{\alpha + d + 1}} + \cdots\, ,
\end{equation}
where we put the summation that correspond to $n_d > \tilde{n}_d$ for each $d$ into dots since, as discussed above, they will not compete for the Newton polygon hence will not alter the analysis with the Puiseux expansion.

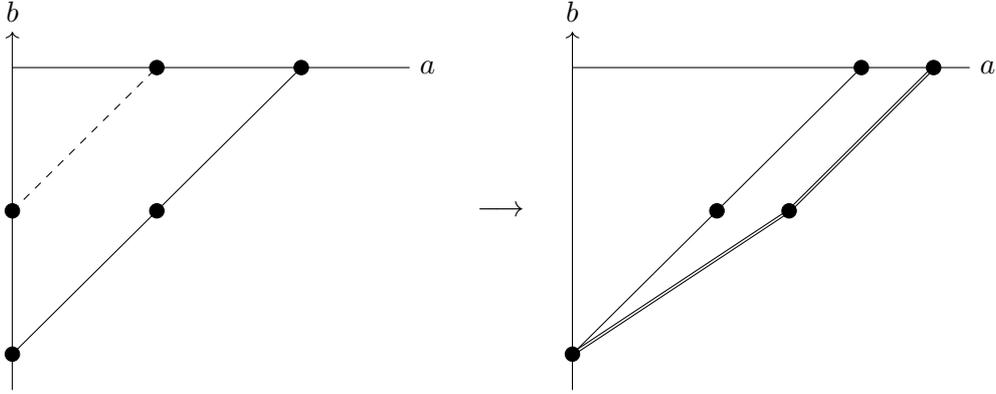
\begin{figure}[H]
  \centering
  \begin{tabular}{ccc}
  \begin{tikzpicture}[scale=.95]
    \draw[->] (0, -4.5) -- (0, 0.5) node[above] {$b$};
    \draw (0, 0) -- (5.5, 0) node[right] {$a$};

    \draw[dashed] (0, -2) -- (2, 0);
    \draw (0, -4) -- (4, 0);

    \draw[fill] (0, -2) circle[radius=0.1];
    \draw[fill] (2, 0) circle[radius=0.1];

    \draw[fill] (0, -4) circle[radius=0.1];
    \draw[fill] (2, -2) circle[radius=0.1];
    \draw[fill] (4, 0) circle[radius=0.1];
  \end{tikzpicture} & \raisebox{14ex}{$\longrightarrow$} &
  \begin{tikzpicture}[scale=.95]
    \draw[->] (0, -4.5) -- (0, 0.5) node[above] {$b$};
    \draw (0, 0) -- (5.5, 0) node[right] {$a$};

    \draw (0, -4) -- (4, 0);
    \draw[double] (0, -4) -- (3, -2) -- (5, 0);
    
    \draw[fill] (3, -2) circle[radius=0.1];
    \draw[fill] (5, 0) circle[radius=0.1];

    \draw[fill] (0, -4) circle[radius=0.1];
    \draw[fill] (2, -2) circle[radius=0.1];
    \draw[fill] (4, 0) circle[radius=0.1];

  \end{tikzpicture}
  \end{tabular}
    \caption{One typical configuration of the Newton diagram generated by a $G_4$-flux containing two different $d$'s. In the left figure, we superpose all Newton diagrams of $\tilde{F}_{d, \tilde{n}_d}$. The solid line corresponds to $\tilde{F}_{\tilde{d}, \tilde{n}_{\tilde{d}}}$ of the highest $\tilde{d}$, and the dashed line corresponds a possible lower $d$. On the right we show the end result. The double lines correspond to the Newton polygon of $\tilde{F}$. In the situation shown in this figure, the two segments of the Newton polygon have both positive slopes, indicating an axion backreaction behavior \eqref{eqn:goodBackreaction} that drives one into the regime of the Distance Conjecture.} \label{fig:NewtonDiagramComponentsGood}
\end{figure}

\begin{figure}[H]
  \centering
  \begin{tabular}{ccc}
  \begin{tikzpicture}[scale=.95]
    \draw[->] (0, -4.5) -- (0, 0.5) node[above] {$\beta$};
    \draw[->] (0, 0) -- (6.5, 0) node[right] {$\alpha$};

    \draw (0, -2) -- (2, 0);
    \draw[dashed] (0, -4) -- (2, -2) -- (4, 0);

    \draw[fill] (0, -2) circle[radius=0.1];
    \draw[fill] (2, 0) circle[radius=0.1];

    \draw[fill] (0, -4) circle[radius=0.1];
    \draw[fill] (2, -2) circle[radius=0.1];
    \draw[fill] (4, 0) circle[radius=0.1];
  \end{tikzpicture} & \raisebox{14ex}{$\longrightarrow$} &
  \begin{tikzpicture}[scale=.95]
    \draw[->] (0, -4.5) -- (0, 0.5) node[above] {$\beta$};
    \draw[->] (0, 0) -- (6.5, 0) node[right] {$\alpha$};

    \draw (0, -2) -- (2, 0);
    \draw[double] (0, -2) -- (2, -4) -- (6, 0);

    \draw[fill] (0, -2) circle[radius=0.1];
    \draw[fill] (2, 0) circle[radius=0.1];

    \draw[fill] (2, -4) circle[radius=0.1];
    \draw[fill] (4, -2) circle[radius=0.1];
    \draw[fill] (6, 0) circle[radius=0.1];
  \end{tikzpicture}
  \end{tabular}
    \caption{Another typical configuration of the Newton diagram generated by a $G_4$-flux containing two different $d$'s. In the left figure, we superpose all Newton diagrams of $\tilde{F}_{d, \tilde{n}_d}$. The solid line corresponds to $\tilde{F}_{\tilde{d}, \tilde{n}_{\tilde{d}}}$ of the highest $\tilde{d}$, and the dashed line corresponds a possible lower $d$. On the right we show the end result. The double lines correspond to the Newton polygon of $\tilde{F}$. In the situation shown in this figure, the segment of the Newton polygon with positive slope will generate an axion backreaction behavior similar to Figure \ref{fig:NewtonDiagramComponentsGood}, while the segment with negative slope will generate a backreaction \eqref{eqn:badBackreaction} that drives $s$ away from the boundary of the complex structure moduli space.} \label{fig:NewtonDiagramComponentsBad}
\end{figure}
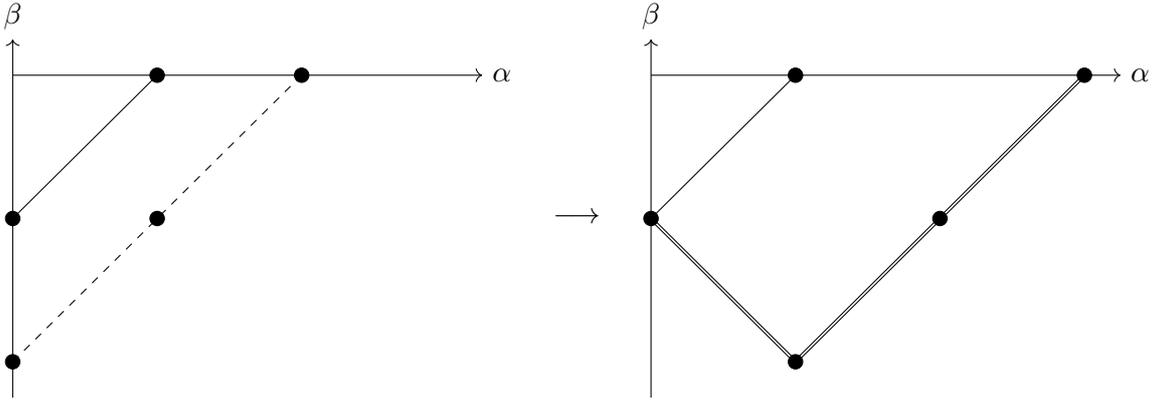

Now we are ready to build the Newton polygon for the polynomial $F$. Firstly we stack all the Newton diagrams of $\tilde{F}_{d, \tilde{n}_d}$ on one $(a, b)$-plane. Generic pictures coming from a $G_4$-flux containing two different $d$'s are shown in the left pictures in Figure \ref{fig:NewtonDiagramComponentsGood} and \ref{fig:NewtonDiagramComponentsBad}. In order to apply the Puiseux expansion, we need to further eliminate all negative powers of $s$ in $F$. Denote the highest $d$ in $G_4$ by $\tilde{d}$, and note that
\begin{equation}
  \tilde{F} = s^{\alpha + \tilde{d} + 1} F = \alpha \, A_{\rm loc} s^{\tilde{d}} + \sum_{d = 0}^4 s^{\tilde{d} - d}\tilde{F}_{d, \tilde{n}_d} + \cdots
\end{equation}
no longer has negative power of $s$ so we can in turn study the solution of $\tilde{F} = 0$ by Puiseux expansions. The extremization condition $F = 0$ is equivalent to $\tilde{F} = 0$ since we assume $s > 0$. 

The resulting Newton diagram of $\tilde{F}$ is simply a combination of the Newton diagrams of various $s^{\tilde{d} - d}\tilde{F}_{d, \tilde{n}_d}$, and each of which is itself the Newton diagram of $\tilde{F}_{d, \tilde{n}_d}$ with a shift towards the $a$-direction, i.e.~along the horizontal axis in Figure \ref{fig:NewtonDiagramComponentsGood} and \ref{fig:NewtonDiagramComponentsBad}, by an amount of $\tilde{d} - d$. In particular, the Newton diagram of $s^{\tilde{d} - d}\tilde{F}_{d, \tilde{n}_d}$ will have dots aligned along a line of slope $+1$ that intersects the $a$-axis at $(\tilde{d} + d - 2\tilde{n}_d, 0)$. Moreover there will be an extra point $(\tilde{d}, 0)$ coming from the term with $A_{\rm loc}$ as coefficient, if $\alpha \ne 0$. The Newton polygon is then found by taking the lower convex hull of the Newton diagram of $\tilde{F}$. Generic pictures of the Newton diagram and Newton polygon of $\tilde{F}$ generated by a $G_4$-flux containing two different $d$'s are shown in the right pictures in Figure \ref{fig:NewtonDiagramComponentsGood} and \ref{fig:NewtonDiagramComponentsBad}.

According to the general structure of the Puiseux expansion reviewed in section \ref{sec:NewtonPuiseux}, each segment in the Newton polygon of $\tilde{F}$ with a positive slope $\gamma > 0$ corresponds to an axion backreaction behavior of the following form
\begin{equation} \label{eqn:goodBackreaction}
  \svev{}(\phi) = c \phi^{\gamma} + \mathcal{O}\left(\frac{1}{\phi}\right)\, ,
\end{equation}
which generalizes the linear backreaction behavior found previously in the literature \cite{Baume:2016psm,Valenzuela:2016yny,Blumenhagen:2017cxt}. Here $c$ is a constant that could depend on the flux numbers. We cannot formulate an general condition on the independence of $c$ on the flux numbers, nevertheless an obvious situation is when there is only a single component $G_4 = g\ket{d, l_d; i_d}$, such that $l_d > 0$. In such simple cases, the flux number $g$ will be factored out in the equation determining $c$, leading to a prefactor $c$ independent from the flux-number $g$.
Note that the analogs of such \textit{sl(2)-elementary fluxes} play a special role in the analysis of the Weak Gravity Conjecture in \cite{Gendler:2020dfp,Bastian:2020egp}.

We also find additional constraints on the possible value of $\gamma$. Recall that our assumption on the value of $\alpha$ is such that it does not make any of the dots in the Newton diagram disappear. This implies that the right-most segment of the Newton polygon will be always of slope $+1$ when it is not formed by translating a single dot, i.e.~it is neither from a component of the form $\ket{d, -d}$ nor from the term containing $A_{\rm loc}$. Note that the convexity condition on the Newton polygon implies that the slope of its various segments increase from left to right. We thus conclude that the backreaction of the large axion on the saxion vev always satisfy $\gamma \le 1$. The $\gamma = 1$ cases exactly agree with the linear backreaction behavior found in \cite{Baume:2016psm,Valenzuela:2016yny,Blumenhagen:2017cxt}. The $0 < \gamma < 1$ cases are more subtle. One can look for these cases by the method that is used study to bad cases discussed below, and we find that the $0 < \gamma < 1$ cases appear exactly when the right-most segment in the Newton polygon is a point. The physical significance of such cases still remains unclear.

Until this point we have assumed that $\alpha$ takes a general value such that none of the coefficients $\cC_{d, n_d; b_d}$ defined in \eqref{eqn:coefficientC} vanishes. Let us relax this assumption and check the consequences. Following the same arguments above, for every $d$ that appears in the flux $G_4$, we need to focus on the highest weight component $\ket{d, \tilde{n}_d; i_d}$ which generates a series of terms in the polynomial $\tilde{F}_{d, \tilde{n}_d}$ that draw a segment of dots in the Newton diagram of $\tilde{F}$. Since we are interested in the lower convex hull in order to obtain the Newton polygon, only missing dots at each end of the segment will likely cause trouble. We see from \eqref{eqn:coefficientC} that for a fixed $\ket{d, \tilde{n}_d; i_d}$, if $\alpha = -d$ then the leftmost dot of the segment disappears, and if $\alpha = d - 2\tilde{n}_d$ then the rightmost dot vanishes. Let us first check what happens if a leftmost dot disappears: Taking into account the condition on the range of $\alpha$ in \eqref{eqn:constraintOnAlpha}, we see that only the dot corresponding to the state $\ket{0, 0}$ could disappear in such a circumstance when $\alpha = 0$. However this will extend the possible values of $\gamma$ and we have also checked that it will not invalidate the discussion of the bad cases in the following subsection, either. On the other hand, when the rightmost dot disappears, new phenomena do appear. We check all possible cases with the method that is used to study bad cases discussed below and find two possible cases that generate backreaction with $\gamma = 2$. They all require $\alpha = 1$ and are given by
\begin{align}
  G_4^{({\rm A})} = g_1\ket{3, 1} + g_2\ket{2, 0} & \quad\Longrightarrow\quad \svev{}_{\rm A}(\phi) = \frac{8g_1^2}{A_{\rm loc} - g_2^2}\phi^2 + \cO\bigg(\frac{1}{\phi}\bigg)\, ,\\
  G_4^{({\rm B})} = h_1\ket{1, 1} + h_2\ket{0, 0} & \quad\Longrightarrow\quad \svev{}_{\rm B}(\phi) = \frac{2h_1^2}{A_{\rm loc} - h_2^2}\phi^2\, ,
\end{align}
where we require that $g_1, h_1 \ne 0$, while $g_2, h_2$ can be switched off. Note that a missing rightmost dot can surely also generate a backreaction with $0 < \gamma < 1$. Note that a backreaction with $\gamma \ne 1$ could potentially generate a polynomial dependence of the backreacted saxion field on the axion travel distance that is in tension with the exponential dependence found in \cite{Baume:2016psm}. This implies that, in the case that our solutions with $\gamma \ne 1$ can be established to exist, our findings are in conflict with the Swampland Distance Conjecture. We stress, however, that we cannot make a conclusive statement about this, since we only focused on the implication of the general asymptotic Hodge theory on the leading contribution in the backreaction of an axion $\phi$ on a single saxion $s$ satisfying the extremization condition ${\partial V}/{\partial s} = 0$. It would be interesting to further investigate the implication of these $\gamma \ne 1$ cases, for example by constructing concrete models realizing them and determine the backreaction effect by imposing minimization conditions.

Having finished the discussion on the cases with $\gamma > 0$, let us look at its contrary: each segment with a negative slope $-\delta < 0$ corresponds to the following solution
\begin{equation} \label{eqn:badBackreaction}
  \svev{}(\phi) = \frac{c}{\phi^\delta} + \mathcal{O}\left(\frac{1}{(\phi)^{\delta + 1}}\right)\, ,
\end{equation}
which implies that the saxion will move away from the boundary of the moduli space as the axion traverse a large field distance. A typical configuration causing such cases is shown in Figure \ref{fig:NewtonDiagramComponentsBad}.
It remains to rule out the backreaction solution of type \eqref{eqn:badBackreaction}. A flux configuration potentially generating solution \eqref{eqn:badBackreaction} is dubbed as a \emph{bad case}. In the next subsection we systematically look for bad cases and show that these solutions \eqref{eqn:badBackreaction} are all invalid.

\subsection{Bad cases and their elimination} \label{app:badCases}

In this section, we systematically analyze bad flux configurations that can potentially generate the following backreaction behavior
\begin{equation} \label{eqn:badCaseDefinition}
  \svev{}(\phi) = \frac{c}{\phi^\delta} + \mathcal{O}\left(\frac{1}{\phi^{\delta + 1}}\right)\, ,
\end{equation}
where $\delta \ge 0$. The Newton polygon of $F$ makes a systematic enumeration of these cases possible.

Let us start by noting that for a root of type \eqref{eqn:badCaseDefinition} to appear, there must be segment of negative slope in the Newton polygon of $F$. Translating this condition to the $G_4$ flux, one sees that there must be at least two different $\tilde{d} > d'$, whose highest weight components correspond to the basis vectors $\ket*{\tilde{d}, \tilde{d} - 2\tilde{n}_{\tilde{d}}}$ and $\ket*{d', d' - 2\tilde{n}_{d'}}$, such that
\begin{equation} \label{eqn:badComponent}
  \boxed{d' - \tilde{n}_{d'} \ge \tilde{d} - \tilde{n}_{\tilde{d}}\, .}
\end{equation}

This instructs us to enumerate the bad cases according to their number of highest weight components, and there are only three possibilities: two, three and four different $d$'s appearing in $G_4$. Using the enhancement rules \cite{Grimm:2018cpv,Grimm:2019ixq}, we can further reduce the possible cases by noting that $d = 4$ and $d = 3$ components cannot co-exist. The same argument also shows that bad cases with four different highest weight components cannot exist, either. Thus we are only listing two- and three-component bad cases in the following subsections.

\subsubsection{Bad cases with two different $d$'s}
These are the fluxes of the following form
\begin{equation}
  G_4 = g_1 \ket*{\tilde{d}, \tilde{d} - 2\tilde{n}_{\tilde{d}}} + g_2 \ket*{d_2, d_2 - 2\tilde{n}_{d_2}}\, ,
\end{equation}
where $\tilde{d} > d_2$ and $\tilde{n}_{\tilde{d}}$ and $\tilde{n}_{d_2}$ satisfy condition \eqref{eqn:badComponent}. There are $16$ possible such fluxes and they are listed in Table \ref{tab:badCasesTwoComponents}.

\begin{table}[H]
\begin{equation*}
\begin{array}{|c|c|c|c|c|c|}\hline
   \textrm{Case}              & g_1         & g_2        & \delta               & c                                                                                                       & \textrm{Reason}\\\hline
  1                           & \ket{4, 0}  & \ket{2, 2} & \multirow{3}{*}{$0$} & \multirow{5}{*}{$\sqrt{-\frac{(\alpha + 4)g_1^2}{(\alpha + 2)g_2^2}}$}                                  & \multirow{7}{*}{\raisebox{-2cm}{Imaginary}}\\\cline{1-3}
  2                           & \ket{4, -2} & \ket{2, 0} &                      &                                                                                                         &\\\cline{1-3}
  3                           & \ket{4, -4} & \ket{2, 0} &                      &                                                                                                         &\\\cline{1-4}
  4                           & \ket{4, -2} & \ket{2, 2} & 1                    &                                                                                                         &\\\cline{1-4}
  5                           & \ket{4, -4} & \ket{2, 2} & 2                    &                                                                                                         &\\\cline{1-5}
  6                           & \ket{4, -2} & \ket{1, 1} & 0                    & \rule[-.4cm]{.0cm}{1.1cm} \sqrt[3]{-\frac{(\alpha + 4)g_1^2}{(\alpha + 1)g_2^2}}                           &\\\cline{1-5}
  \rule[-.2cm]{.0cm}{.55cm} 7 & \ket{3, -1} & \ket{1, 1} & 0                    & \multirow{2}{*}{\raisebox{-0.3cm}{$\sqrt{-\frac{(\alpha + 3)g_1^2}{(\alpha + 1)g_2^2}}$}}            &\\\cline{1-4}
  \rule[-.2cm]{.0cm}{.55cm} 8 & \ket{3, -3} & \ket{1, 1} & 1                    &                                                                                                         &\\\cline{1-5}
  9                           & \ket{4, -4} & \ket{1, 1} & \frac{2}{3}          & \rule[-.4cm]{.0cm}{1.1cm} \sqrt[3]{-\frac{(\alpha + 4)g_1^2}{(\alpha + 1)g_2^2}} (1 , \omega, \omega^2) &\\\hline
  10                          & \ket{3, 1}  & \ket{2, 2} & \multirow{2}{*}{$0$} & \multirow{5}{*}{$-\frac{(\alpha + 3)g_1^2}{(\alpha + 2)g_2^2}$}                                         & \multirow{7}{*}{Negative}\\\cline{1-3}
  11                          & \ket{3, -1} & \ket{2, 0} &                      &                                                                                                         &\\\cline{1-4}
  12                          & \ket{3, -1} & \ket{2, 2} & \multirow{2}{*}{$2$} &                                                                                                         &\\\cline{1-3}
  13                          & \ket{3, -3} & \ket{2, 0} &                      &                                                                                                         &\\\cline{1-4}
  14                          & \ket{3, -3} & \ket{2, 2} & 4                    &                                                                                                         &\\\cline{1-5}
  15                          & \ket{2, 0}  & \ket{1, 1} & 0                    & \multirow{2}{*}{$-\frac{(\alpha + 2)g_1^2}{(\alpha + 1)g_2^2}$}                                         &\\\cline{1-4}
  16                          & \ket{2, -2} & \ket{1, 1} & 2                    &                                                                                                         &\\\hline
\end{array}
\end{equation*}
\caption{Bad cases with two different $d$'s. In the $c$ column we have omitted non-essential normalization factors to display their shared properties. The last column points out the reason that invalidates the solution to $c$. Case 9 has three solutions to its prefactor $c$, and they differ from each other by a factor of $\omega = e^{2 \pi \im / 3}$.} \label{tab:badCasesTwoComponents}
\end{table}

We hereby present an analysis of case 13. We denote its flux configuration as
\begin{equation} \label{eqn:exampleFlux}
  G_4 = g_1 \ket{3, -3} + g_2 \ket{2, 0}\, .
\end{equation}
Such a flux configuration can appear, for example, in a degeneration of Calabi-Yau fourfold with $h^{3, 1} = 3$. Following the notation of the singularity types in \cite{Grimm:2019ixq}, the enhancement of singularity type can be $\rII_{0, 1} \to \rV_{3, 3}$.

The asymptotic scalar potential reads
\begin{equation} \label{eqn:examplePotential}
  V(s, \phi) = \frac{1}{s^\alpha} \bigg( \frac{\hg_1^2}{s^3} + \hg_2^2 + \frac{2 \hg_2^2 \phi^2}{s^2} - A_{\textrm{loc}} \bigg)\, .
\end{equation}
The extremization condition is given by
\begin{equation} \label{ean:exampleFtilde}
  0 = \tilde{F}(s, \hphi) = \alpha (A_{\textrm{loc}} - \hg_2^2) s^3 - 4(\alpha + 2)\hg_2^2 \hphi^{-2}s - (\alpha + 3)\hg_1^2\, .
\end{equation}

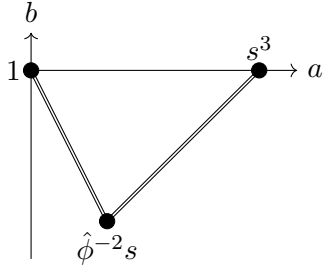
\begin{figure}[H]
  \centering
  \begin{tikzpicture}[scale=1]
    \draw[->] (0, -2.5) -- (0, 0.5) node[above] {$b$};
    \draw[->] (0, 0) -- (3.5, 0) node[right] {$a$};

    \draw[double] (0, 0) -- (1, -2);
    \draw[double] (1, -2) -- (3, 0);

    \draw[fill] (0, 0) circle[radius=0.1] node[left] {$1$};
    \draw[fill] (1, -2) circle[radius=0.1] node[below] {$\hphi^{-2}s$};
    \draw[fill] (3, 0) circle[radius=0.1] node[above] {$s^3$};
  \end{tikzpicture}
  \caption{The Newton polygon of \eqref{ean:exampleFtilde}. It has a segment generating backreaction $\svev{} \sim c \phi$ and another segment corresponding to $\svev{} \sim c \phi^{-2}$.} \label{fig:exampleNewton2}
\end{figure}

Upon inspecting the scalar potential \eqref{eqn:examplePotential}, a necessary condition for it to be not blowing up when $s \to \infty$ is $\alpha \ge 0$. Furthermore, we impose $A_{\textrm{loc}} > 0$, otherwise there will be a runaway towards $s \to \infty$. We display the Newton polygon in Figure \ref{fig:exampleNewton2}.

From the Newton polygon in Figure \ref{fig:exampleNewton2}, we read out that there are two possible solutions, one with leading term proportional to $\phi$ and the other with leading term proportional to $1/\phi^2$. More explicitly, they are
\begin{align}
  \svev{}_{1, \pm}(\phi) & = \pm \sqrt{\frac{4(\alpha + 2)\hg_2^2}{\alpha (A_{\textrm{loc}} - \hg_2^2)}} \phi + \cO\bigg(\frac{1}{\phi}\bigg)\, ,\nonumber\\
  \svev{}_2(\phi)        & = -\frac{(\alpha + 3)\hg_1^2}{4(\alpha + 2)\hg_2^2}\frac{1}{\phi^2} + \cO\bigg(\frac{1}{\phi^3}\bigg)\, .
\end{align}

Note that among these three roots, only $\svev{}_{1, +}$ is positive when $A_{\rm loc} > \hg_2^2$. We thus conclude that for the flux \eqref{eqn:exampleFlux}, either there is a runaway in $s$, or there is a vacuum with the linear backreaction behavior
\begin{equation}
  \svev{}(\phi) = \sqrt{\frac{4(\alpha + 2)\hg_2^2}{\alpha (A_{\textrm{loc}} - \hg_2^2)}} \phi + \cO\bigg(\frac{1}{\phi}\bigg)\, ,
\end{equation}
when the axion $\phi$ is large. We would also like to point out the $\hg_2$ is actually a product between $g_2$ and a non-vanishing function on the saxions other than $s$. Depending on the value of these saxions, the linearly backreacted $s$ could even disappear. Another remark is that the leading coefficient in $\svev{}(\phi)$ depends on the flux number $g_2$ and the localized contribution $A_{\rm loc}$, indicating that the backreaction effect could be delayed. A further investigation into such cases is left for future work.

\subsubsection{Bad cases with three different $d$'s}
\begin{figure}[H]
  \centering
  \begin{tabular}{ccc}
  \begin{tikzpicture}
    \draw[->] (0, -2.5) -- (0, 0.5) node[above] {$\beta$};
    \draw[->] (0, 0) -- (5.5, 0) node[right] {$\alpha$};

    \draw[double] (0, -2) -- (2, -2) -- (3, -2) -- (5, 0);
    
    \draw[fill] (0, -2) circle[radius=0.1] node[left] {$-2$};
    \draw[fill] (2, 0) circle[radius=0.1] node[above] {$2$};
    \draw[fill] (2, -2) circle[radius=0.1];
    \draw[fill] (3, -2) circle[radius=0.1];
    \draw[fill] (4, 0) circle[radius=0.1] node[above] {$4$};
    \draw[fill] (5, 0) circle[radius=0.1] node[above] {$5$};
  \end{tikzpicture} & &
  \begin{tikzpicture}
    \draw[->] (0, -2.5) -- (0, 0.5) node[above] {$\beta$};
    \draw[->] (0, 0) -- (4.5, 0) node[right] {$\alpha$};

    \draw[double] (0, -2) -- (1, -2) -- (2, -2) -- (4, 0);

    \draw[fill] (0, -2) circle[radius=0.1] node[left] {$-2$};
    \draw[fill] (1, -2) circle[radius=0.1];
    \draw[fill] (2, 0) circle[radius=0.1] node[above] {$2$};
    \draw[fill] (2, -2) circle[radius=0.1];
    \draw[fill] (3, 0) circle[radius=0.1] node[above] {$3$};
    \draw[fill] (4, 0) circle[radius=0.1] node[above] {$4$};
  \end{tikzpicture}
  \end{tabular}
  \caption{The left picture is the Newton diagram of equation \eqref{eqn:F1}, and the right picture corresponds to \eqref{eqn:F2}. Note the both diagrams show a possible linear backreaction behavior in addition to the bad constant backreaction solution.} \label{fig:bad3}
\end{figure}
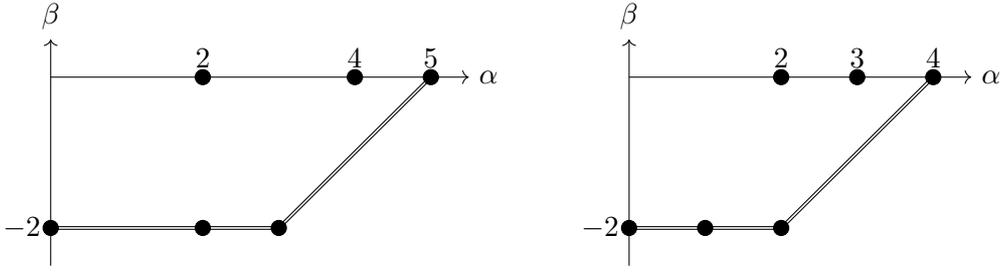

In this subsection, we list all \emph{essential} flux configurations containing three different $d$'s that are likely bad. By essential, we mean that the focus will be on the cases whose leading backreaction coefficient is determined by all three components. In the situation where this coefficient is only determined by two components, it reduces to one of the cases listed in Table \ref{tab:badCasesTwoComponents}. Upon inspecting possible shapes of the Newton diagram, one sees that we need to find a flux configuration
\begin{equation}
  G_4 = g_1\ket*{\tilde{d}, \tilde{d} - 2\tilde{n}_{\tilde{d}}} + g_2 \ket{d_2, d_2 - 2\tilde{n}_{d_2}} + g_3 \ket{d_3, d_3 - 2\tilde{n}_{d_3}}\, ,
\end{equation}
satisfying the following conditions
\begin{align}
  \tilde{d}                          & > d_2 \ne d_3\, ,\\
  \tilde{d} - \tilde{n}_{\tilde{d}}  & = d_2 - \tilde{n}_{d_2} = d_3 - \tilde{n}_{d_3}\, ,\\
  \tilde{d} - 2\tilde{n}_{\tilde{d}} & < d_2 - 2\tilde{n}_{d_2} < d_3 - 2\tilde{n}_{d_3}\, .
\end{align}
In the end there are only two bad cases and we discuss them in turn. The first one is given by
\begin{equation}
  G_4^{(1)} = g_1\ket{4, -2} + g_2\ket{2, 0} + g_3\ket{1, 1}\, ,
\end{equation}
which induces a scalar potential of the form
\begin{equation}
  V_1(s, \phi) = \frac{1}{s^\alpha}\bigg( \frac{16 \hg_1^2 \phi^2}{s^4} + \frac{\hg_1^2}{s^2} + \hg_2^2 + \frac{4\hg_2^2 \phi^2}{s^2} + \frac{\hg_3^2 \phi^2}{s} + \hg_3^2 s - A_{\rm loc} \bigg).
\end{equation}
The condition one imposes on $\alpha$ is that $\alpha \ge 1$. The extremization condition $\tilde{F}_1 = 0$ is given by
\begin{equation} \label{eqn:F1}
  \tilde{F}_1 = (1 - \alpha)\hg_3^2 s^5 + \alpha(A_{\rm loc} - \hg_2^2)s^4 - (1 + \alpha)\hg_3^2 \hphi^{-2} s^3 - (2 + \alpha)(\hg_1^2 + 4\hg_2^2 \hphi^{-2})s^2 - 16(4 + \alpha)\hg_1^2 \hphi^{-2}\, ,
\end{equation}
whose Newton diagram is given in the left picture in Figure \ref{fig:bad3}. From the diagram we see that there is a potential bad root with $\delta_1 = 0$. The pre-factor $c_1$ should satisfy the following cubic equation
\begin{equation} \label{eqn:quartic}
  (1 + \alpha) g_3^2 c_2^3 + 4g_2^2 (2 + \alpha) c_2^2 + 16(4 + \alpha)g_1^2 = 0\, .
\end{equation}
Note that the coefficient in every term in the above equation is positive. In other words, there is no sign-flip in the list of coefficients in the above polynomial equation with real coefficients. The Descartes' rule of signs tells us that the number of positive root of a real polynomial equation is bounded by the number of sign-flips in its list of coefficients. Hence we conclude that even if the above quartic equation has a real root $c_2$, it will nevertheless be negative. This rules out the first bad case.

The last case to consider is given by
\begin{equation}
  G_4^{(2)} = g_1\ket{3, -1} + g_2\ket{2, 0} + g_3\ket{1, 1}\, .
\end{equation}
The corresponding scalar potential has the form
\begin{equation}
  V_2(s, \phi) = \frac{1}{s^\alpha}\bigg( \frac{9 \hg_1^2 \phi^2}{s^3} + \frac{\hg_1^2}{s} + \hg_2^2 + \frac{4\hg_2^2 \phi^2}{s^2} + \frac{\hg_3^2 \phi^2}{s} + \hg_3^2 s - A_{\rm loc} \bigg).
\end{equation}
And one has again the constraint $\alpha \ge 1$. Its extremization condition $\tilde{F}_2 = 0$ is given by
\begin{equation} \label{eqn:F2}
  \tilde{F}_2 = (1 - \alpha)\hg_3^2 s^4 + \alpha(A_{\rm loc} - \hg_2^2)s^3 - (1 + \alpha)(\hg_1^2 + \hg_3^2 \hphi^{-2}) s^2 - 4(2 + \alpha)\hg_2^2 \hphi^{-2} s^2 - 9(3 + \alpha)\hg_1^2 \hphi^{-2}\, ,
\end{equation}
whose Newton diagram is given in the right picture in Figure \ref{fig:bad3}. It indicates again a potential bad root with $\delta_2 = 0$, whose pre-factor $c_2$ satisfies the following quadratic equation
\begin{equation}
  \hg_3^2 c_2^2 + 4(2 + \alpha) \hg_2^2 c_2 + 9(3 + \alpha) \hg_1^2 = 0\, ,
\end{equation}
which has no positive real root again by Descartes' rule of sign. Hence this case is ruled out.

To conclude, we have ruled out all possible bad cases which induce the backreaction behavior \eqref{eqn:badCaseDefinition} by showing that their accompanying pre-factor $c$ is either negative or purely imaginary. This leads to our conclusion that a large displacement of an axion $\phi$ can only backreacts on its saxion partner $s$ in the way shown in equation \eqref{eqn:goodBackreaction} with rational exponent $0 < \gamma \le 2$.

\subsubsection*{Acknowledgments}

It is a great pleasure to thank Brice Bastian, Damian van de Heisteeg, Eran Palti, Erik Plauschinn, and
especially Irene Valenzuela for very useful discussions and correspondence. 
This research 
is partly supported by the Dutch Research Council (NWO) via a Start-Up grant and a Vici grant.

\appendix
\section{More detailed properties of the commuting $\slt$-triples} \label{app:commutingsl2}

This appendix fills in some detail of the derivation of \eqref{eqn:exe-1} and \eqref{eqn:limitingPropertiesWithAde}. We follow the proof of lemma (4.5) in \cite{MR1042802}. In order to do this we need to take a closer look at the limiting mixed Hodge structure and the induced splittings on the infinitesimal isometry Lie algebra $\fg$. For definiteness we will (mostly) align with the mathematical notations in \cite{MR1042802} in this appendix.

Recall that the Lie algebra $\gR$ consists of the infinitesimal isometries for weight four variation of Hodge structure. Concretely, it can be identified with $\gR = \mathfrak{so}(2 + h^{2, 2}_{\rm p}, 2h^{1, 3})$, where $h^{p, q}$ are the Hodge numbers of the family of Calabi-Yau fourfolds and $h^{2, 2}_{\rm p}$ is the complex dimension of the space of primitive $(2, 2)$-forms. The complexification of $\gR$ is denoted by $\fg$ and can be identified with $\fg = \mathfrak{so}(2 + h^{2,2}_{\rm p} + 2h^{1, 3}, \bbC)$. We present these identifications merely to make the exposition more concrete but these will not be used in the following discussion.

Recall further that the log-monodromy operators $N_1, \ldots, N_n$ introduced in \eqref{eqn:defnLogMonodromy} define the \emph{monodromy weight filtration} $W^{(n)} = W(N_1 + \cdots + N_n)[-4]$ on the primitive middle cohomology $\Hp^4(Y, \bbC)$. The monodromy weight filtration together with the limiting Hodge filtration $\Fpol$ introduced in \eqref{eqn:nilpotentOrbit} define the \emph{limiting mixed Hodge structure} $(\Fpol, W^{(n)})$ on $\Hp^4(Y, \bbC)$. We denote the \emph{Deligne splitting} associated to the $(\Fpol, W^{(n)})$ by
\begin{equation}
  \Hp^4(Y, \bbC) = \bigoplus_{p, q} I^{p, q}\, ,
\end{equation}
satisfying
\begin{equation} \label{eqn:propertyDeligneSplitting}
  \Fpol^p = \bigoplus_{r \ge p} I^{r, s}\, ,\quad W^{(n)}_k = \bigoplus_{r + s \le k} I^{r, s}\, ,\quad \textrm{ and } I^{p, q} = \conj{I^{q, p}} \mod \bigoplus_{\substack{r < p\\s < q}} I^{r, s}\, .
\end{equation}
The Deligne splitting is functorial, which puts a Deligne splitting on the Lie algebra
\begin{equation} \label{eqn:gDeligneSplitting}
  \fg = \bigoplus_{p, q} \fg^{p, q}\, ,
\end{equation}
whose components can be concretely identified as
\begin{equation} \label{eqn:characterisationOfGPQ}
  \fg^{p, q} = \left\{ X \in \fg \mid X(I^{r, s}) \subs I^{r + p, s + q}\right\}\, .
\end{equation}

Recall that the $\SLt$-orbit theorem constructs a set of commuting $\slt$-triples, whose lowering and number operators are denoted $N_r^-$ and $N_r^0$, respectively in section \ref{sec:commSl2}. We have also defined the partial sums
\begin{equation}
  \opY{r} = N_1^0 + \cdots + N_r^0\, , \quad \textrm{ for all } r\, .
\end{equation}
Moreover, the $\SLt$-orbit theorem constructs a series of $\bbR$-split special mixed Hodge structures from the limiting mixed Hodge structure $(\Fpol, W^{(n)})$. We denote these mixed Hodge structures by $(F_{(r)}, W^{(r)})$ and briefly state their relation to the nilpotent orbit data $(\Fpol, N_1, \ldots, N_n)$. Firstly, there is a Hodge filtration $F_{(n)}$ built out of the data $(\Fpol, N_1, \ldots, N_n)$. The $\bbR$-split mixed Hodge structure $(F_{(n)}, W^{(n)})$ is called the $\SLt$-splitting of the limiting mixed Hodge structure $(\Fpol, W^{(n)})$. Then each of the remaining $(F_{(r)}, W^{(r)})$ is built recursively by taking the $\SLt$-splitting of the mixed Hodge structure $(e^{\im N_{r + 1}}, W^{(r + 1)})$, where the weight filtration is give by $W^{(r)} = W(N_1 + \cdots + N_r)[-4]$. The construction of $\SLt$-splitting is a bit involved and we refer to the math papers \cite{MR840721} and \cite{MR1042802} for more precise information. This is also reviewed recently in a physics paper \cite{Grimm:2018cpv}.

We will need three important properties of the mixed Hodge structures $(F_{(r)}, W^{(r)})$. Firstly, although the weight filtrations $W^{(r)}$ are defined using the operators $N_1, \ldots, N_r$, it turned out (by $\SLt$-orbit theorem) that it agrees with the monodromy weight filtration defined by the lowering operators in the commuting $\slt$-triples
\begin{equation}
  W^{(r)} = W(N_1 + \cdots + N_r)[-4] = W(N_1^- + \cdots + N_r^-)[-4]\, .
\end{equation}
Secondly, denote the Deligne splitting of $(F_{(r)}, W^{(r)})$ by
\begin{equation}
  \Hp^4(Y, \bbC) = \bigoplus_{p, q} I^{p, q}_{(r)}\, .
\end{equation}
We have
\begin{equation} \label{eqn:-1-1morphism}
  N_{(r)}(I^{p, q}_{(r)}) \subs I^{p - 1, q - 1}_{(r)}\, .
\end{equation}
Lastly, the eigenspaces of the number operators $\opY{r}$ are defined in terms of $I^{p, q}_{(r)}$ as
\begin{equation} \label{eqn:YIsGradingOfMHS}
  \opY{r}v = l v\, ,\quad \textrm{ for all } v \in \bigoplus_{r + s = l + 4} I^{r, s}_{(r)}\, .
\end{equation}

There is another splitting of the real Lie algebra $\gR$ coming from the commuting $\slt$-triples. Since the number operators commute with each other, their partial sums also mutually commute
\begin{equation}
  [\opY{r}, \opY{s}] = 0\, , \quad \textrm{for all } r, s\, .
\end{equation}
Using the Jacobi identity, one sees that the adjoint actions $\ad{\opY{r}}(\blank) = [\opY{r}, \blank]$ on the Lie algebra $\fg$ also commute with each other
\begin{equation}
  [\ad{\opY{r}}, \ad{\opY{s}}] = 0\, , \quad \textrm{for all } r, s\, .
\end{equation}
Hence these commuting adjoint actions induce a multi-grading on the Lie algebra $\fg$
\begin{equation} \label{eqn:gMultiGrading}
  \fg = \bigoplus_{\Bell = (l_1, \ldots, l_n)} \fg_{\Bell}\, ,
\end{equation}
where each component is the simultaneous eigenspace of all $\ad{\opY{r}}$:
\begin{equation}
  [\opY{r}, X] = l_r X\, , \quad \textrm{for all } X \in \fg_{\Bell} \textrm{ and } r = 1, \ldots, n\, .
\end{equation}
The eigenvalues $l_1, \ldots, l_r$ are all integers due to the general property of $\slt$-representations.

These two splittings \eqref{eqn:gDeligneSplitting} and \eqref{eqn:gMultiGrading} will be the central objects in this appendix.

\subsection{The map $e(s) e^{\phi^i N_i} e(s)^{-1}$}
Let us reproduce the definition of the $e(s)$-operator for convenience
\begin{equation} \label{eqn:defnEOperatorInSAgain}
  e( s^1, \ldots, s^n ) = \exp\left\{ \hf ( \log s^r ) N^0_r \right\}\, ,
\end{equation}
where in the exponential we sum over $r$. It turns out that rewriting the above definition in terms of the partial sums $\opY{r}$ is more suitable for our purpose in this appendix. To achieve this we formally set $s^{n + 1} = 1$ and redefine variables
\begin{equation}
  \sigma^r = \frac{s^r}{s^{r + 1}}\, ,\quad \textrm{ for all } r = 1, \ldots, n\, .
\end{equation}
With the variables $\sigma^r$ one has
\begin{equation} \label{eqn:defnEOperatorInSigma}
  e( s^1, \ldots, s^n ) = \exp\left\{ \hf ( \log \sigma^r ) \opY{r} \right\}\, .
\end{equation}
We will proceed using the new form \eqref{eqn:defnEOperatorInSigma} of the $e(s)$-operator.

Let us spell out the expression $e(s) e^{\phi^i N_i} e(s)^{-1}$ we want to compute
\begin{align}
  e(s) e^{\phi^i N_i} e(s)^{-1} & = \sum_{k = 1}^\infty \frac{(\phi^i)^k}{k!} e(s) N_i^k e(s)^{-1}\nonumber\\
                                & = \sum_{k = 1}^\infty \frac{(\phi^i)^k}{k!} (\operatorname{Ad}_{e(s)}(N_i))^k\nonumber\\
                                & = e^{\phi^i \operatorname{Ad}_{e(s)}(N_i)}\, ,
\end{align}
where $\operatorname{Ad}_{e(s)}(N_i) = e(s) N_i e(s)^{-1}$ is the adjoint  action of the group element $e(s)$ on the Lie algebra element $N_i$. Recall that the adjoint action of the Lie algebra on itself is defined to satisfy
\begin{equation}
  \operatorname{Ad}_{e^X}(Y) = e^{\ad{X}}(Y)\, ,\quad \textrm{ for all } X, Y \in \fg\, .
\end{equation}
So we have reduced the quantity that we want to compute into
\begin{equation} \label{eqn:intermediateAppendixA1}
  e(s) e^{\phi^i N_i} e(s)^{-1} = \exp\left\{\phi^i e^{\hf (\log \sigma^r) \ad{\opY{r}}}(N_i)\right\}\, ,
\end{equation}
where summation over $i$ and $r$ are assumed and we have used the new form \eqref{eqn:defnEOperatorInSigma} of $e(s)$-operator. From this expression we see that the important data one needs is the commutator $[\opY{r}, N_i]$. This commutator depends on the relative position of $r$ and $i$.

Let us first consider the case when $r \ge i$. From the property \eqref{eqn:-1-1morphism}, we see (by contradiction) that the same property must hold for each $N_i$ with $i \le r$
\begin{equation} \label{eqn:Ni-1-1Morphism}
  N_{i}(I^{p, q}_{(r)}) \subs I^{p - 1, q - 1}_{(r)}\, ,\quad \textrm{ for } i = 1, \ldots, r\, .
\end{equation}
This automatically forces the relation
\begin{equation}
  [\opY{r}, N_i] = -2 N_i\, ,\quad \text{for all } r \ge i\, ,
\end{equation}
by the characterization \eqref{eqn:characterisationOfGPQ} of the Deligne splitting on the Lie algebra and the property \eqref{eqn:YIsGradingOfMHS}.

Next we look at the case $r < i$. In such case we no longer have good control over the commutator $[\opY{r}, N_i]$. The best result one has is simply that $N_i$ preserves the filtration $W^{(r)}$. One way to see this is to use an explicit characterization of the monodromy weight filtration in remark (2.3) of \cite{MR791673} with the $(-4)$-shift
\begin{equation}
  W^{(r)}_l = \sum_{j \ge \max\{-1, l - 4\}} (\Ker N^{j + 1}_{(r)}) \cap (\Im N^{j - l + 4}_{(r)})\, ,
\end{equation}
and note that $N_i$ commutes with $N_{(r)}$.

Using again \eqref{eqn:characterisationOfGPQ}, \eqref{eqn:YIsGradingOfMHS}, and the general property of the Deligne splitting \eqref{eqn:propertyDeligneSplitting}, we conclude that the eigen-decomposition of $N_i$ with respect to the action of $\adY{r}$ has only \emph{non-positive} eigenvalues.

We can have more control over the eigenvalues by investigating the multi-grading \eqref{eqn:gMultiGrading}. Remembering that $r < i$, let us diagonalize actions of all $\adY{r}$ on $N_i$ simultaneously. We split $N_i = \hN_i + N_i'$, such that $\adY{r}(\hN_i) = 0$ for all $r = 1, \ldots, i - 1$. And further decompose the remaining $N_i'$ according to the multi-grading \eqref{eqn:gMultiGrading}
\begin{equation}
  N_i' = \sum_{\alpha^i_1, \ldots, \alpha^i_{i - 1} > 0} N'_{i, \alpha^i_1, \ldots, \alpha^i_{i - 1}},
\end{equation}
where $\alpha^i_r > 0$ labels the eigenvalue of $\adY{r}$ on $N_i$,
\begin{equation}
  \Big[\opY{r}, N'_{i, \alpha^i_1, \ldots, \alpha^i_{i - 1}}\Big] = - \alpha^i_r N'_{i, \alpha^i_1, \ldots, \alpha^i_{i - 1}}\, ,\quad \text{for } i > r.
\end{equation}
Moreover, it can be checked that each $\alpha^i_r$ is also an integer, and the components $\hN_i$, $N'_{i, \ua^i}$ are nilpotent.

To summarize, for every $i = 1, \ldots, n$, one has a decomposition
\begin{equation}
  N_i = \hN_i + \sum_{\ua^i > 0} N'_{i, \ua^i}\, ,
\end{equation}
where $\ua^i = (\alpha^i_1, \ldots, \alpha^i_{i - 1})$ denotes the collection of positive integer eigenvalues, such that $[\opY{r}, \hN_{i}] = 0$ for all $r > i$, and
\begin{equation} \label{eqn:adYOnN}
  [\opY{r}, N_i] =
  \begin{cases}
    -2N_i\, ,                                      & \text{for } i \le r\, ,\\
    -\sum_{\ua^i > 0} \alpha^i_r N'_{i, \ua^j}\, , & \text{for } i > r\, .
  \end{cases}
\end{equation}

It is then straightforward to plug \eqref{eqn:adYOnN} into \eqref{eqn:intermediateAppendixA1} and conclude
\begin{equation}
  e(s) e^{ \phi^i N_i } e(s)^{-1} = \exp\left\{ \sum_{i = 1}^n \frac{\phi^i}{s^i} \left( \hN_i + \sum_{\ua^i > 0} \frac{N'_{i, \ua^i}}{(\frac{s^1}{s^2})^{\alpha^i_{1}/2} \cdots (\frac{s^{i - 1}}{s^i})^{\alpha^i_{i - 1}/2}} \right) \right\}\, .
\end{equation}
This finishes the derivation of equation \eqref{eqn:exe-1}. We would like to point out that due to the nilpotent operators, this equation is actually \emph{polynomial} in $\phi$ and $s$. This is in contrast to the quantity we want to compute in the next subsection.

\subsection{The map $e(s) e^{\Gamma(z)} e(s)^{-1}$ in the limit}

Following the procedure in the previous subsection, we have
\begin{equation} \label{eqn:intermediateAppendixA2}
  e(s) e^{\Gamma(z)} e(s)^{-1} = \exp\left\{e^{\hf (\log \sigma^r) \ad{\opY{r}}}(\Gamma(z))\right\} = 1 + e^{\hf (\log \sigma^r) \ad{\opY{r}}}(\Gamma(z)) + \cdots\, ,
\end{equation}
where a summation over $r$ under the exponential is assumed. In the above expression we only displayed up to the first order term in the outer exponential in hindsight as it will turn out that this first order term will be \emph{exponentially suppressed} in the limit $\sigma^r \to \infty$, so that the expression \eqref{eqn:intermediateAppendixA2} will approach $1$ exponentially as shown in the first equation in \eqref{eqn:limitingPropertiesWithAde}.

The expression \eqref{eqn:intermediateAppendixA2} again instructs us to look into the commutator $[\opY{r}, \Gamma(z)]$. Unfortunately, we cannot work out an expression for \eqref{eqn:intermediateAppendixA2} as ``concrete'' as \eqref{eqn:exe-1}. The best general result we have here is the limit.

Let us first state some general property of the mapping $\Gamma(z)$ following \cite{MR1042802}. Firstly, the mapping $\Gamma(z)$ is holomorphic in $z$ and satisfies $\Gamma(0) = 0$, which means that it enjoys a series expansion around $z = 0$
\begin{equation} \label{eqn:holomorphicGamma}
  \Gamma(z_1, \ldots, z_n) = \sum_{k_1, \ldots, k_n \ge 0} \Gamma^{k_1, \ldots, k_n} z_1^{k_1} \cdots z_n^{k_n}\, ,
\end{equation}
with $\Gamma^{\underline{0}} = 0$.
Secondly, proposition (2.6) in \cite{MR1042802} states that $\Gamma(z)$ satisfies
\begin{equation}
  [N_j, \Gamma(z_1, \ldots, z_j = 0, \ldots, z_n)] = 0\, ,\quad \textrm{for all } j = 1, \ldots, n\, .
\end{equation}
Combining the above identity for all $j \le r$, we have
\begin{equation} \label{eqn:crucialPropertyOfGamma1}
  [N_{(r)}, \Gamma(0, \ldots, 0, z_{r + 1}, \ldots, z_n)] = 0\, ,\quad \textrm{for all } r = 1, \ldots, n\, .
\end{equation}

Let us decompose the series expansion of $\Gamma(z)$ in \eqref{eqn:holomorphicGamma} further with respect to the multi-grading \eqref{eqn:gMultiGrading}
\begin{equation}
  \Gamma^{k_1, \ldots, k_n}  = \sum_{\Bell} \Gamma^{k_1, \ldots, k_n}_{\Bell}\, .
\end{equation}
The conclusion here is that, for a fixed set of $\Bell = (l_1, \ldots, l_n)$, if a component $l_r > 0$, then
\begin{equation} \label{eqn:crucialPropertyOfGamma2}
  \Gamma^{0, \ldots, 0, k_{r + 1}, \ldots, k_n}_{l_1, \ldots, l_r > 0, \ldots, l_n} = 0\, ,\quad \textrm{ for all }  k_{r + 1}, \ldots, k_n\, .
\end{equation}
This can be seen, e.g., again by looking at the general expression of the weight filtration on the Lie algebra $\fg$. Note that the monodromy weight filtration on the Lie algebra $\fg$ no longer has the $(-4)$-shift, so one concludes that operators commute with $N_{(r)}$ lives below level $l_r \le 0$.

Define
\begin{equation}
  \Gamma_{\Bell}(z) = \sum_{k_1, \ldots, k_n} \Gamma^{k_1, \ldots, k_n}_{\Bell}\, .
\end{equation}
Let us check the first order term in \eqref{eqn:intermediateAppendixA2}. For a fixed $\Bell = (l_1, \ldots, l_n)$, we have
\begin{equation} \label{eqn:intermediateAdjointYOnGammaL}
  e^{\hf (\log \sigma^r) \ad{\opY{r}}}(\Gamma_{\Bell}(z)) = (\sigma^1)^{\frac{l_1}{2}} \cdots (\sigma^n)^{\frac{l_n}{2}} \Gamma_{\Bell}(z)\, .
\end{equation}

We would like to find the limit $e(s) e^{\Gamma(z)} e(s)^{-1}$ as $\sigma^r \to \infty$ for all $r$. In order to do so, we choose any norm $\norm{\blank}$ on the Lie algebra $\fg$ and first check $\norm{\Gamma_{\Bell}(z)}$. For all possible $j = 1, \ldots, n$, there are two possibilities: If $l_j > 0$, then with \eqref{eqn:crucialPropertyOfGamma2}, we have
\begin{equation}
  \norm{\Gamma_{l_1, \ldots, l_j > 0, \ldots, l_n}(z)} = \norm{\sum_{k_1, \ldots, k_j > 0} \Gamma^{k_1, \ldots, k_n}_{l_1, \ldots, l_n} z_1^{k_1} \cdots z_n^{k_n}} \le M \sum_{i = 1}^j e^{-c_i s^i}\, ,
\end{equation}
for some positive constants $M$ and $c_i$ in the limit $\sigma^r \to \infty$ for all $r$. We have used the relation $z_i = e^{2\pi\im t^i} = e^{-2\pi s^i} e^{2\pi\im \phi^i}$. Plug this back into \eqref{eqn:intermediateAdjointYOnGammaL} and we have
\begin{equation} \label{eqn:crucialEstimate}
  \norm{e^{\hf (\log \sigma^r) \ad{\opY{r}}}(\Gamma_{\Bell}(z))} \le M (\sigma^1)^{\frac{l_1}{2}} \cdots (\sigma^n)^{\frac{l_n}{2}} \sum_{i = 1}^j e^{-c_i \sigma^i \cdots \sigma^n}\, .
\end{equation}
Note that we have used the relation $s^j = \sigma^j \cdots \sigma^n$.

A second possibility is $l_j \le 0$ and in such cases the estimate \eqref{eqn:crucialEstimate} holds trivially (recall that $\Gamma_{\Bell}(0) = 0$ for all $\Bell$).

In conclusion, the first order term in \eqref{eqn:intermediateAppendixA2} satisfies the estimate \eqref{eqn:crucialEstimate}, which means that in the limit $\sigma^r \to \infty$ it goes to 0 exponentially. We have thus conclude that $e(s) e^{\Gamma(z)} e(s)^{-1} \to 1$ with exponentially suppressed corrections in the limit $\sigma^r \to \infty$ for all $r$.

\subsection{The filtration $e(s) F_0$ in the limit}
We will be short in this section and mainly refer the reader to the papers \cite{MR840721} and \cite{MR1042802}. The result we would like to show is that in the limit where all $\sigma^r \to \infty$, one has
\begin{equation}
  e(s) F_0 \to F_{(n)}\, .
\end{equation}
Combining with the definition that
\begin{equation}
  F_\infty = e^{\im \hN_{(n)}}F_{(n)}\, ,
\end{equation}
this shows the second equation of \eqref{eqn:limitingPropertiesWithAde}.

Here are some facts \cite{MR1042802} about the relation between $F_0$ and $F_{(n)}$. There exists an operator $\eta \in \gR$ such that
\begin{equation}
  F_0 = e^\eta F_{(n)}\, .
\end{equation}
The operator $\eta$ satisfies
\begin{equation} \label{eqn:etaInLambda-1-1}
  \eta (I^{p, q}) \subs \bigoplus_{\substack{r < p\\r < q}} I^{r, s}\, .
\end{equation}
Moreover the operator $\eta$ commutes with every $(r, r)$-morphism of the mixed Hodge structure $(W^{(n)}, F_{(n)})$.

Let us decompose the operator $\eta$ with respect to the multi-grading \eqref{eqn:gMultiGrading}
\begin{equation}
  \eta = \sum_{\Bell} \eta_{\Bell}\, .
\end{equation}
Then the property \eqref{eqn:etaInLambda-1-1} implies that $l_n < 0$. Furthermore, recall from \eqref{eqn:Ni-1-1Morphism} every $N_r$ with $r = 1, \ldots, n - 1$ is a $(-1, -1)$-morphism of the mixed Hodge structure $(W^{(n)}, F_{(n)})$, we have $[N_r, \eta] = 0$ for all $r = 1, \ldots, n - 1$. According to the definition of the monodromy weight filtration, this implies that $l_r \le 0$. So we have

\begin{align}
  e(s) e^{\eta} e(s)^{-1} & = \exp\left\{ e^{\hf (\log \sigma^r) \ad{\opY{r}}} (\eta) \right\} \nonumber\\
                          & = \exp\left\{ \sum_{\substack{l_1, \ldots, l_{n - 1} \le 0\\l_n < 0}} (\sigma^1)^{\frac{l_1}{2}} \cdots (\sigma^n)^{\frac{l_n}{2}} \eta_{l_1, \ldots, l_n} \right\} \to 1\, ,
\end{align}
in the limit where every $\sigma^r \to \infty$.

\bibliographystyle{jhep}
\bibliography{bibfile}

\end{document}